\documentclass[a4paper,fleqn]{cas-dc}
\usepackage{url}
\usepackage[numbers,sort]{natbib}
\usepackage{academicons}
\usepackage{pifont}
\usepackage{longtable}
\usepackage{multirow}
\usepackage[utf8]{inputenc}
\usepackage{float}
\usepackage{caption}
\usepackage{enumitem}
\usepackage{booktabs}
\usepackage[most]{tcolorbox}
\usepackage{siunitx}
\usepackage[framemethod=TikZ]{mdframed}
\tcbuselibrary{listingsutf8}
\usepackage{lipsum}
\usepackage{subcaption}
\usepackage{booktabs}             
\usepackage{tabularx}   
\usepackage{subcaption}
\usepackage{array}      
\usepackage{adjustbox}
\usepackage{rotating} 
\usepackage{ragged2e}  
\usepackage{cellspace}
\usepackage{adjustbox}
\usepackage{placeins}
\usepackage{lineno}
\usepackage[normalem]{ulem}
\usepackage{xparse}

\tcbuselibrary{listings, breakable}
\newcommand{\Gherkin}[1]{\color{Green}\texttt{\textbf{#1}}\color{Black}\xspace}

\newcolumntype{S}[1]{>{\hspace{5pt}}p{#1}<{\hspace{5pt}}}
\definecolor{headerblue}{RGB}{0, 81, 158}
\definecolor{countorange}{RGB}{204, 102, 0}
\newtcblisting{promptbox}{
  listing engine=minted,
  colback=blue!5!white,
  colframe=blue!75!black,
  listing only,
  minted language=text,
  minted options={
    frame=none,
    linenos=true,
    firstnumber=1,
    breaklines=true,
    fontsize=\footnotesize
  },
  left=5pt,
  right=5pt,
  top=5pt,
  bottom=5pt,
  enhanced,
  sharp corners
}
\newtcolorbox{promptboxplain}{
  colback=blue!5!white,
  colframe=blue!75!black,
  sharp corners,
  boxrule=0.5pt,
  fontupper=\footnotesize\ttfamily,
  enhanced,
  left=5pt,
  right=5pt,
  top=5pt,
  bottom=5pt
}
\definecolor{redbg}{rgb}{1.0, 0.95, 0.95}
\tcbset{
  redbox/.style={
    breakable,
    enhanced,
    colback=redbg,
    colframe=black,
    boxrule=0.15pt,
    arc=3pt,
    top=1pt,
    bottom=2pt,
    left=4pt,
    right=4pt
  }
}
\definecolor{greenbg}{rgb}{0.95, 1.0, 0.95}
\tcbset{
  greenbox/.style={
    breakable,
    enhanced,
    colback=greenbg,
    colframe=black,
    boxrule=0.15pt,
    arc=3pt,
    top=2pt,
    bottom=2pt,
    left=4pt,
    right=4pt
  }
}
\definecolor{light-gray}{gray}{0.9}
\tcbset{
  myframe/.style={
    breakable,
    enhanced,
    colback=light-gray,
    colframe=black,
    boxrule=0.15pt,
    arc=3pt,
    top=2pt,
    bottom=2pt,
    left=4pt,
    right=4pt
  }
}
\def\tsc#1{\csdef{#1}{\textsc{\lowercase{#1}}\xspace}}
\tsc{WGM}
\tsc{QE}
\tsc{EP}
\tsc{PMS}
\tsc{BEC}
\tsc{DE}

\makeatletter
\DeclareRobustCommand{\mathcolor}[2]{%
  \begingroup
  \ifmmode
    \textcolor{#1}{#2}%
  \else
    \textcolor{#1}{#2}%
  \fi
  \endgroup
}
\makeatother

\newcommand\change[3]{\linelabel{#2}\textcolor{blue}{#1} \comm{C\ref{#3}} \linelabel{#2end}  }

\newcounter{commentnumber}

\renewcommand{\change}[3]{#1\xspace}

\begin{document}

\shorttitle{}

\let\WriteBookmarks\relax
\def\floatpagepagefraction{1}
\def\textpagefraction{.001}

\title[mode = title]{From Law to Gherkin: A Human-Centred Quasi-Experiment on the Quality of LLM-Generated Behavioural Specifications from Food-Safety Regulations}

\ExplSyntaxOn
\cs_set:Npn \__first_footerline:
  {
    \group_begin:
      \small\sffamily
      {\mbox{}}
    \group_end:
  }
\ExplSyntaxOff
\shorttitle{}

\author[1]{Shabnam Hassani}[orcid=0009-0008-3056-4073]  
\author[1]{Mehrdad Sabetzadeh}[orcid=0000-0002-4711-8319]
\author[1]{Daniel Amyot}[orcid=0000-0003-2414-1791]

\affiliation[1]{organization={School of EECS},
                addressline={University of Ottawa}, 
                city={Ottawa},
                country={Canada}}

\cortext[cor1]{Corresponding author: \href{mailto:m.sabetzadeh@uottawa.ca}{m.sabetzadeh@uottawa.ca}}

\begin{abstract}
\noindent\textbf{Context:} Laws and regulations increasingly influence software design, development, and quality assurance in regulated domains; however, the technology-neutral formulation of legal provisions complicates the derivation of concrete specifications, requirements, and acceptance criteria needed to verify software compliance. Producing these artifacts manually is labour-intensive and error-prone. Recent advances in generative AI, particularly large language models (LLMs), offer the potential for automated assistance in deriving software engineering artifacts from legal texts.

\noindent\textbf{Objective:} Following a quasi-experimental design, we present the first systematic human-subject evaluation of \textcolor{black}{LLMs' ability to automatically derive Gherkin behavioural specifications from legal texts. Gherkin is a domain-specific language for specifying system behaviours through scenario-based descriptions written in the \texttt{Given–When–Then} format.
Due to their structured and machine-readable nature, Gherkin specifications lend themselves more readily to automation within software-development processes.}

\noindent\textbf{Methods:} We recruited 10 participants to evaluate Gherkin specifications generated from food-safety regulations by two LLMs, Claude and Llama. Sixty specifications were generated. Each participant independently assessed 12 specifications across five quality criteria: \emph{relevance}, \emph{clarity}, \emph{completeness}, \emph{singularity}, and \emph{time savings}. Each specification was evaluated by two participants, yielding 120 assessments with quantitative ratings and qualitative feedback.

\noindent\textbf{Results:} \textcolor{black}{Ratings were uniformly high (top-two categories): relevance 95\%, clarity 100\%, completeness 94.2\%, singularity 93.4\%, and time savings 91.7\%. No statistically reliable differences were observed across participants or between LLMs. Qualitative feedback noted occasional omissions, hallucinations, and mixed intents; the first two, in particular, underscore the importance of human oversight, especially in safety-critical domains where non-compliance can have severe consequences.}

\noindent\textbf{Conclusion:} \textcolor{black}{Our results suggest that, in the context of food safety, LLMs can assist in deriving Gherkin specifications from legal texts; however, observed omissions and hallucinations necessitate systematic human review.}
\end{abstract}

\begin{keywords}
Legal Compliance Automation \sep
Large Language Models (LLMs) \sep Behaviour-Driven Development \sep Gherkin \sep Quasi-Experiment
\end{keywords}

\maketitle

\section{Introduction} \label{sec:intro}
Laws and regulations increasingly shape software engineering practices, affecting design, development, and assurance in regulated domains. Legal provisions impose a variety of constraints, concerning, among others, safety, security, and privacy. Non-compliance can lead to undue risks, financial penalties, and reputational damage. Consequently, inaccuracies in translating legal provisions into software engineering artifacts can be highly costly.

Legal texts are commonly phrased in technology-agnostic language to maintain broad applicability and long-term relevance. This often complicates their application in concrete technological contexts~\cite{cordella2024regulating,Hassani2025Empirical}.
Translating legal texts into software engineering artifacts, such as behavioural specifications or acceptance criteria, remains a significant challenge. Practitioners often have difficulty interpreting legal texts in a way that leads to precise, verifiable requirements and acceptance criteria necessary to demonstrate compliance~\cite{Breaux2008Analyzing,maxwell2011legal}. 
Manual translation is time-consuming, error-prone, and inconsistent, increasing the risk of non-compliance, especially in domains where regulatory adherence is critical to public safety~\cite{elahidoost2024practices,kosenkov2024developing,angermeir2024towards}.
Existing research in requirements engineering (RE) has primarily tackled legal compliance through methods of requirements extraction, classification, and ambiguity resolution from legal provisions~\cite{Breaux2006Towards,Zeni2015Gaiust,Zeni2016Building,abualhaija2020automated,Torre2020Ai,Hey2020NoRBERT,Deshpande2021BERT,Chatterjee2021Pipeline,Amaral2022AI,marques2024using,Hassani2025Empirical}.
Yet, little attention has been given to systematically transforming legal texts into structured, machine-analyzable software artifacts such as behavioural specifications.

\subsection{Motivation}\label{sec:motovation}
Recent advances in Generative AI (GenAI), particularly Large Language Models (LLMs), present a promising opportunity to directly derive behavioural specifications from legal provisions. 
Early efforts have shown that Natural Language Processing (NLP) techniques can be used to generate software engineering artifacts -- such as acceptance tests, test cases, code stubs, and control flow graphs (test paths) -- from stakeholder-written specifications, natural-language requirements, and  use-case descriptions~\cite{soeken2012assisted, lafi2021automated, fischbach2023automatic}. These approaches mostly depend on rule-based and heuristic NLP methods, graph traversal, and pattern matching. More recently, researchers have begun using LLMs to generate software engineering artifacts, particularly test cases. Mathur et al.~\cite{mathur2023automated} employ prompting to derive unit tests from requirements. Ferreira et al.~\cite{Ferreira2025Acceptance} generate Gherkin scenarios and test cases from user stories. Alagarsamy et al.~\cite{alagarsamy2024enhancing} fine-tune LLMs to produce syntactically correct Java unit tests  from method-level descriptions. Although these studies advance automated test generation in software engineering, they do not address the application of LLMs to legal texts.

\textcolor{black}{In this article, we present the first human-subject study that systematically evaluates the ability of two state-of-the-art LLMs -- Llama and Claude -- to generate behavioural specifications from legal texts. We focus specifically on \emph{Gherkin}~\cite{GherkinSpecification}, a domain-specific language for expressing system behaviours through scenario-based descriptions written in the \texttt{Given–When–Then} format. Gherkin's structured, semi-formal syntax bridges natural language and executable specifications, making it particularly well suited for assessing whether LLMs can produce precise representations of legal provisions. Because Gherkin scenarios remain close to everyday language while enforcing an explicit behavioural structure, they can be understood and validated by stakeholders from diverse backgrounds, supporting collaboration and transparency in the RE process~\cite{salgueiro2021best}. At the same time, Gherkin specifications are machine-readable and widely used in agile and behaviour-driven development (BDD) methods~\cite{wanderley2015evaluation,dos2018automated,salgueiro2021best}, which allows them to be readily integrated into automated software engineering workflows.}

Our study is framed using Basili's Goal-Question-Metric (GQM) paradigm~\cite{basili1994goal}. The overarching goal is to evaluate how well Llama and Claude can generate Gherkin specifications from legal texts, with the aim of making legal texts more accessible to software engineers. The concrete research questions and evaluation metrics used to operationalize this goal are detailed in Section~\ref{sec:RQs}.

We address a specific instance of the broader challenge of translating legal texts into software artifacts, focusing on the domain of \emph{\bfseries food safety}. We choose this domain for three main reasons: First, food safety is a high-impact domain affecting all individuals~\cite{WHO2024FoodSafety} and representing a substantial global market~\cite{FoodSafetyTesting}. \change{Second, food safety is increasingly intertwined with Industry~4.0 and IoT-based monitoring systems, leading to software-relevant legal requirements; however, because food-safety regulations are written in technology-neutral language, a gap remains between legal provisions and the software systems required to implement them~\cite{Bouzembrak2019Internet,dadhaneeya2023internet}.}{r1c2}{reviewer1Comment2} Third, food safety offers the advantage of allowing us to build on our prior experience and a previously developed conceptual model of software-relevant concepts in this domain~\cite{Hassani2025Empirical}, supporting deeper reflection on the challenges and the formulation of precise guidelines for our study.

\subsection{Illustrative Example}
\begin{figure*}
  \centering
  \begin{subfigure}[b]{0.48\linewidth}
    \includegraphics[width=\linewidth]{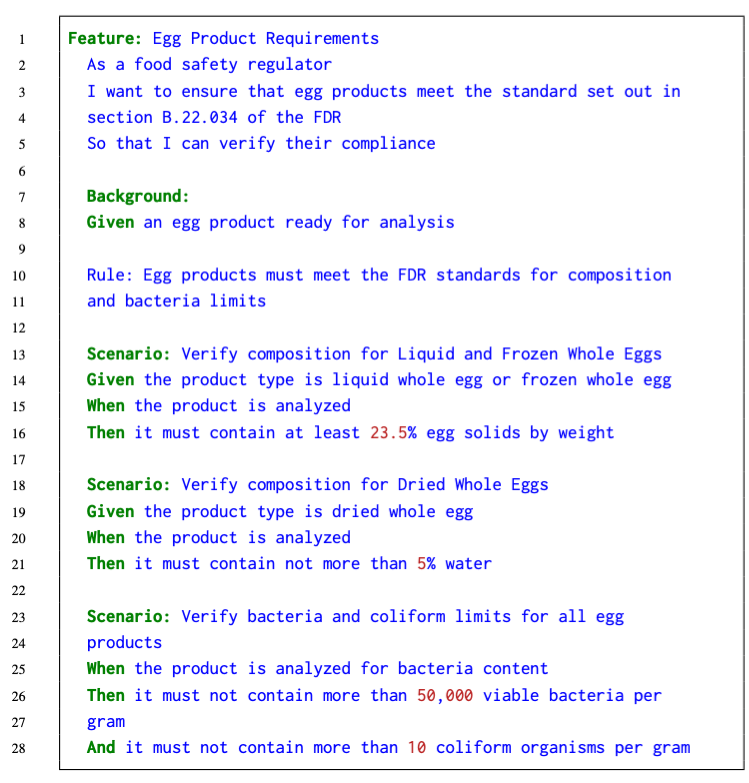}
\caption{Generated by Claude 3.7 Sonnet: each scenario addresses a distinct requirement for different egg types.}

  \end{subfigure}
  \hfill
  \begin{subfigure}[b]{0.48\linewidth}
    \includegraphics[width=\linewidth]{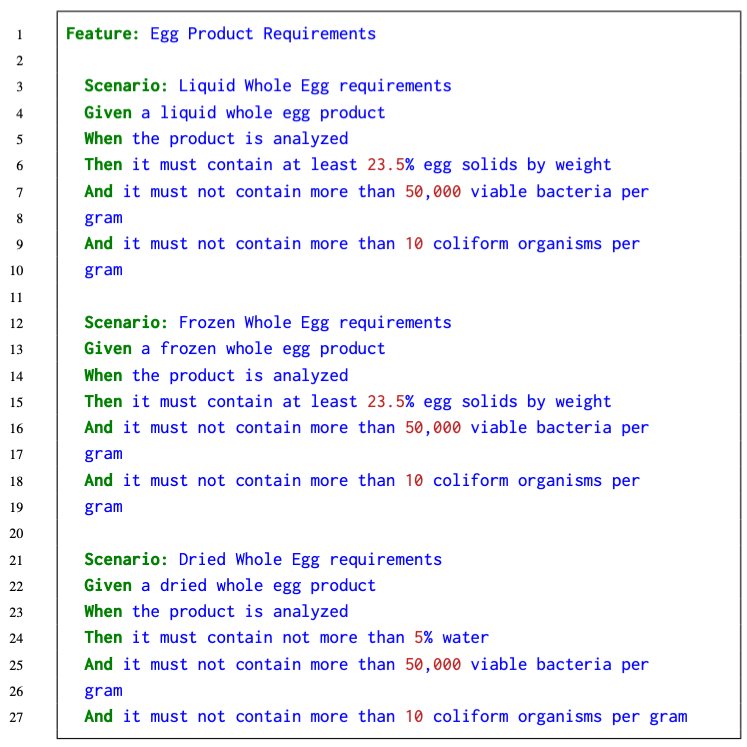}
 \caption{Generated by Llama 3.3 70B Instruct: each scenario combines composition and microbiological checks, violating \emph{singularity} and reducing \emph{clarity} of individual objectives.}
  \end{subfigure}
  \caption{Example Gherkin specifications generated by two different LLMs for the same food-safety legal provision: (a)~illustrates singular and focused scenarios; ~(b) illustrates mixed-objective scenarios.}
  \label{fig:gherkin-subfigs}
  \vspace*{-1em}
\end{figure*}

Figure~\ref{fig:gherkin-subfigs} shows an example Gherkin specification, based on a legal provision incorporated by reference into the Safe Food for Canadians Regulations (SFCR)~\cite{SFCR2018}\footnote{\url{https://bit.ly/4o4tEDq}}, and generated using Claude~3.7~Sonnet~\cite{Claude} and Llama~3.3~70B~Instruct~\cite{Llama}. The provision is as follows:

\begin{tcolorbox}[myframe]
Liquid Whole Egg, Dried Whole Egg and Frozen Whole Egg are the foods that meet the standard set out in section B.22.034 of the FDR and
\begin{enumerate}[label=(\alph*), leftmargin=2em]
    \item in the case of liquid whole egg and frozen whole egg, contain at least 23.5\% egg solids by weight;
    \item in the case of dried whole egg, contain not more than 5\% water; and
    \item contain not more than
    \begin{enumerate}[label=(\roman*), leftmargin=3em]
        \item 50,000 viable bacteria per gram; and
        \item 10 coliform organisms per gram.
    \end{enumerate}
\end{enumerate}
\end{tcolorbox}

Different LLMs may generate Gherkin specifications that vary in quality across different criteria. For instance, the Gherkin specification shown in Figure~\ref{fig:gherkin-subfigs}(a) is generated by Claude 3.7 Sonnet~\cite{Claude}. It follows BDD best practices by presenting scenarios that each focus on a single, well-defined objective. In contrast, the specification in Figure~\ref{fig:gherkin-subfigs}(b), generated by Llama 3.3 70B Instruct~\cite{Llama}, blends multiple purposes within individual scenarios, such as 1)~measuring product weight, 2)~assessing water content, and 3)~evaluating microorganism levels. BDD scenarios are generally more effective when each scenario addresses only one distinct goal.
Our objective in this article is to evaluate, through a quasi-experimental study, the quality of Gherkin specifications generated by LLMs from food-safety regulations, and to identify any recurring issues in such specifications.

\subsection{Research Questions (RQs)}\label{sec:RQs}
To operationalize our GQM goal (Section~\ref{sec:motovation}), we answer the following RQs:\\[.3em]
\textbf{RQ1:} \textit{How well do LLMs perform in generating Gherkin specifications from \change{food-safety}{}{reviewer1Comment1}legal provisions?}\\
\textbf{RQ2:} \textit{How do Llama and Claude compare when generating Gherkin specifications from \change{food-safety}{}{reviewer1Comment1} legal provisions?}\\[-.5em]

We answer RQ1 and RQ2 by evaluating the quality of LLM-generated Gherkin specifications along five criteria: 
\emph{relevance}, the degree to which the specification matches the intended system behaviour as described in the legal text;
\emph{completeness}, the degree to which all functions and characteristics implied by the legal requirement are included without omissions; 
\emph{clarity}, the degree to which the specification is written in a clear and unambiguous manner, allowing unique interpretation; 
\emph{singularity}, the degree to which scenarios focus on one purpose, avoiding mixed-purpose scenarios. \change{A singular scenario addresses a distinct, atomic objective. This contrasts with ``mixed-purpose'' scenarios, which conflate multiple independent requirements (e.g., verifying both physical composition and microbial limits) into a single test flow}{r2c2}{reviewer2Comment1}; 
and, \emph{time savings}, the degree to which the generated specification can be reused or adapted, thus reducing manual authoring effort. \change{We note that, in our study, \emph{time savings} is based on a self-reported judgment of efficiency, not on direct time-tracking.}{r2c1}{reviewer2Comment1}
Each criterion is rated on a five‐point scale. 
\change{Detailed definitions and the full rating scales used by participants are provided in Table~\ref{tab:evaluation-criteria} (Section~\ref{sec:GherkinEvaluation}).}{r2c1-1}{reviewer2Comment1}

\section{Background}\label{sec:Back}
This section presents the necessary background on food-safety regulations and the enabling technologies used.

\subsection{Food-Safety Regulations}
Countries worldwide maintain regulatory frameworks to ensure food-supply safety~\cite{WHO2024FoodSafety}. In Canada, the Safe Food for Canadians Regulations (SFCR)~\cite{SFCR2018} consolidate federal requirements for importing, exporting, and inter-provincial food trade, with product-specific obligations deferred to the Food-Specific Requirements and Guidance (FSRG)~\cite{FSRG2018} covering domains such as dairy, meat, fish, and produce. Our evaluation uses the SFCR and FSRG texts and, where explicitly cross-referenced, related instruments including the Safe Food for Canadians Act (SFCA) and the Food and Drug Regulations (FDR) under the Food and Drugs Act (FDA).

We build on our prior work~\cite{Hassani2025Empirical}, where we introduced a taxonomy of software-relevant concepts in food-safety regulations and released a labelled dataset of such provisions from Canadian and U.S. regulations. In this study, we use a subset of 30 provisions from that dataset~\cite{Data}.

\subsection{Behaviour-Driven Development and Gherkin}
Behaviour-driven development (BDD) is a collaborative software engineering method for expressing requirements as concrete, human-readable behaviour scenarios that are accessible to both technical and non-technical stakeholders~\cite{BDDbestpractice}. We use established quality metrics (Section~\ref{sec:BDDquality}) to evaluate scenarios generated from legal texts using LLMs.

To represent these scenarios, we employ Gherkin, a domain-specific language commonly used in BDD. Gherkin provides a structured \texttt{Given-When-Then} template: \Gherkin{Given} defines the context, \Gherkin{When} describes the event or action, and \Gherkin{Then} asserts the expected outcome~\cite{GherkinSpecification}. Gherkin can be mapped to executable tests via step definitions; tools such as JBehave~\cite{jbehaveOfficial} and Cucumber~\cite{hellesoy2017cucumber} match Gherkin feature files with automation code~\cite{GherkinSpecification,hellesoy2017cucumber}. While we do not implement such automation here, our study lays the groundwork for such integration in legal compliance. Gherkin feature files are organized using a small set of keywords -- \Gherkin{Feature}, \Gherkin{Background}, \Gherkin{Scenario}, \Gherkin{Given}, \Gherkin{When}, \Gherkin{Then}, \Gherkin{And}, \Gherkin{But}, \Gherkin{Scenario Outline}, and \Gherkin{Examples}~\cite{GherkinSpecification} -- and may include doc strings (\Gherkin{"""}), data tables (\Gherkin{{|}}), tags (\Gherkin{{@}}), and comments (\Gherkin{{\#}})~\cite{hellesoy2017cucumber}. Figure~\ref{fig:gherkin-subfigs} (Section~\ref{sec:intro}) shows examples using these constructs.

\change{In our study, we view LLM-derived scenarios from food-safety regulations (e.g., those in Figure~\ref{fig:gherkin-subfigs}) as black-box specifications for a nominal software system. As also stated earlier, the regulations are technology-neutral, so the scenarios do not assume a particular implementation. Instead, they are analogous to a system-level interaction model: they capture exchanges between external actors and the proposed system while omitting internal design details (which cannot be inferred from the regulations). Put differently, the scenarios state how a compliant system must respond to external triggers, thus operationalizing the regulations into testable system-boundary interactions.
}{r1c8}{reviewer1Comment8}

\section{Methodology}\label{sec:approach}
This section presents the research process we adopt in order to investigate the effectiveness of using LLMs to derive Gherkin specifications from food-safety legal provisions.

Our study has three main phases: (1)~deriving Gherkin specifications using LLMs, (2)~evaluating the resulting specifications through participant ratings and feedback, and (3)~analyzing the collected responses both quantitatively and qualitatively. 
This process involves recruiting participants, generating specifications, conducting human evaluations, and synthesizing input from participants.%
\footnote{Ethics approval for our study was granted by the University of Ottawa's Research Ethics Board (file number \hbox{H-08-24-10663}).}

For our study design, we concluded that a fully controlled experiment was infeasible under our real-world constraints, as it would require rigid standardization of numerous potential confounders -- time of day, participant fatigue, and ambient environmental conditions -- that we simply could not enforce given our participants' varied schedules. Moreover, we deliberately recruited participants whose experience ranged from senior undergraduates to final-year PhD students in order to preserve external validity; narrowing this spectrum would have both limited recruitment and reduced the relevance of our findings. Taken together, the logistical challenge of strict control over extraneous factors and the need to accommodate a diverse cohort would have significantly undermined the internal validity of a fully controlled experiment.

Similarly, a randomized crossover design~\cite{montgomery2017design}, where each participant is exposed to all treatments in a random order, was ruled out since we wanted (1)~every Gherkin specification to be reviewed by two participants, and (2)~each participant to review six specifications generated by Llama and six by Claude. These constraints made it infeasible to randomly assign treatments while ensuring balanced task coverage. Therefore, our treatment assignment was driven by study-design constraints and participant-task balancing, rather than full randomization.

Because we cannot control all key experimental variables, our study employs a \emph{quasi-experimental} design.
As described by Jedlitschka et al.~\cite{jedlitschka2008reporting}, a quasi-experiment is an empirical inquiry similar to a controlled experiment, except that treatment assignment is determined by subject or object characteristics rather than full random allocation. 
In our case, while task selection for each participant involves random sampling to ensure task coverage without overlap, the overall assignment is guided by study-design requirements, specifically balanced exposure to both models (Claude and Llama), while also ensuring that no participant evaluated Gherkin specifications for the same legal provision from both models.

\subsection{Participant Recruitment}
The first step is participant selection. Eligibility is determined by the following: 

1) \textit{Technical Background}: A)~Proficiency with BDD principles and Gherkin syntax. B)~Experience in programming, writing and reviewing test cases. C)~Completion of specific university-level courses such as ``Software Requirement Analysis'' and ``Software Quality Assurance'' ensuring foundational knowledge of requirements and test specification; 
2) \textit{Language Proficiency}: Ability to comprehend and produce written work in English, as study materials and instructions are provided in English; 
and 3) \textit{Open Eligibility}: In alignment with Equity, Diversity, and Inclusion (EDI) principles, no restrictions were applied based on race, sex, or geographic location. However, since we wanted to compensate participants, we focused on the pool of students at the University of Ottawa.

To verify eligibility, we asked interested participants to send us a short email describing their background. Specifically, they were instructed to: 1)~indicate whether they had completed the relevant courses and when; 2)~describe their familiarity with BDD and Gherkin syntax; 3)~summarize their experience with writing test cases and reviewing software requirements. Prospective participants were also asked to attach a copy of their CV to support their self-reported qualifications. We reviewed this information to ensure that all selected participants met the minimum technical criteria.

Participants were recruited through multiple channels in the software engineering and computer science programs at the University of Ottawa. The recruitment strategies included:
1)~\textit{In-class presentations}: Short presentations in advanced software engineering courses explained the study scope and voluntary nature of participation. 2)~\textit{Post-class flyers}: Flyers were distributed with detailed study information and instructions for potential volunteers on how to sign up~\cite{Data}; and 3)~\textit{Emails and announcements}: Targeted emails were sent to eligible students in relevant courses, detailing the tasks and compensation structure.

Participants were included on a first-come, first-served basis, subject to eligibility, to ensure fairness.
When more candidates responded than needed (target sample size: 10), those not selected received follow-up emails thanking them for their interest and explaining that recruitment had closed.

Participants were compensated for their time and effort. The compensation structure was as follows: 1)~\textit{Per task}: \$8 for each completed Gherkin evaluation task. 2)~\textit{Completion bonus}: An additional \$54 for completing all 12 tasks. 3)~\textit{Total compensation}: Participants could earn up to \$150. This tiered structure ensured participants could withdraw at any time while still receiving proportional compensation for completed tasks.

All participants were required to review and sign a consent form~\cite{Data}, detailing the study goals, procedures, and the right to withdraw. 
The consent process ensured that participation was informed, voluntary, and conducted with respect to ethical standards, including anonymity. Data collection for the study was conducted between February 2025 and March 2025.

\change{Given that all participants were university students, we took steps to mitigate potential power dynamics in researcher-participant relationships. The first author, who had no instructional or evaluative role with respect to participants, handled recruitment and data collection and served as the sole point of contact; the second and third authors had no direct interaction with participants.}{r3c4}{reviewer3Comment4}

\change{As a final remark regarding participant selection, our choice to recruit software engineering students is consistent with established empirical guidelines. Prior work, including from Salman et al.~\cite{Salman2015students}, indicates that such students can serve as suitable proxies for professionals in experiments involving comprehension, structured quality assessment, and artifact evaluation -- tasks that closely match those in our study.}{r1c3-1}{reviewer1Comment3}

\subsection{LLM-based Gherkin Specification Generation} \label{sec:Generation}

In the second step, we used an LLM -- either Llama or Claude -- to derive Gherkin specifications from food-safety legal provisions. \change{This step was carried out via the respective chat interfaces for these models, namely the Hugging Face chat interface for Llama and the Claude web interface. As these interfaces do not expose model parameters to the user, all generations were performed using the default settings.}{r1c7}{reviewer1Comment7}\change{}{}{reviewer3Comment5}

\begin{figure} 
  \centering
  \includegraphics[width=\linewidth]{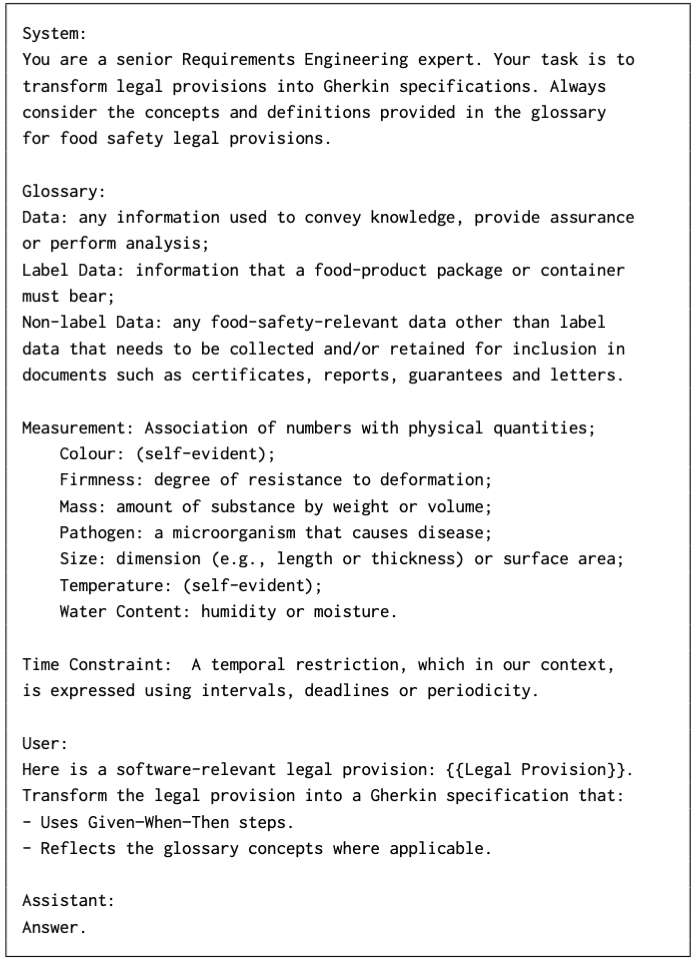}
    \caption{Prompt for generating a Gherkin specification.}\label{fig3:prompt}
    \vspace{-1em}
\end{figure}

Figure~\ref{fig3:prompt} shows our prompt template. \change{To improve the consistency, task alignment, and interpretability of LLM outputs, our prompt design follows established prompt-engineering principles: (i)~\emph{clear and specific task definition}, (ii)~\emph{role assignment to guide the model's perspective}~\cite{wu2023large,kong2024better}, and (iii)~\emph{contextual grounding through domain-specific information}~\cite{openai_prompt_best_practices,palantir_prompt_best_practices,anthropic_prompt_best_practices}.
}{r3c5}{reviewer3Comment5}

The prompt has three main components: a glossary of domain-specific concepts, a food-safety legal provision, and a task instruction. 
The glossary is drawn from our prior work~\cite{Hassani2025Empirical}, which defines concepts such as \emph{Data}, \emph{Measurement}, and \emph{Time Constraint} to support consistent interpretation of software-relevant food-safety provisions. 
The provisions are sampled from data released alongside our prior work~\cite{Hassani2025Empirical}. 
The task instruction asks the LLM to produce a behavioural specification in the Gherkin syntax, aligning each provision with relevant glossary concepts and structuring the output in the \texttt{Given-When-Then} format. 

\subsection{Participant Evaluation of the Resulting Gherkin Specifications} \label{sec:GherkinEvaluation}

\begin{table*}
  \centering
 \caption{Definitions and corresponding scales for the quality criteria evaluated by participants.}
   \label{tab:evaluation-criteria}
  \includegraphics[width=\linewidth]{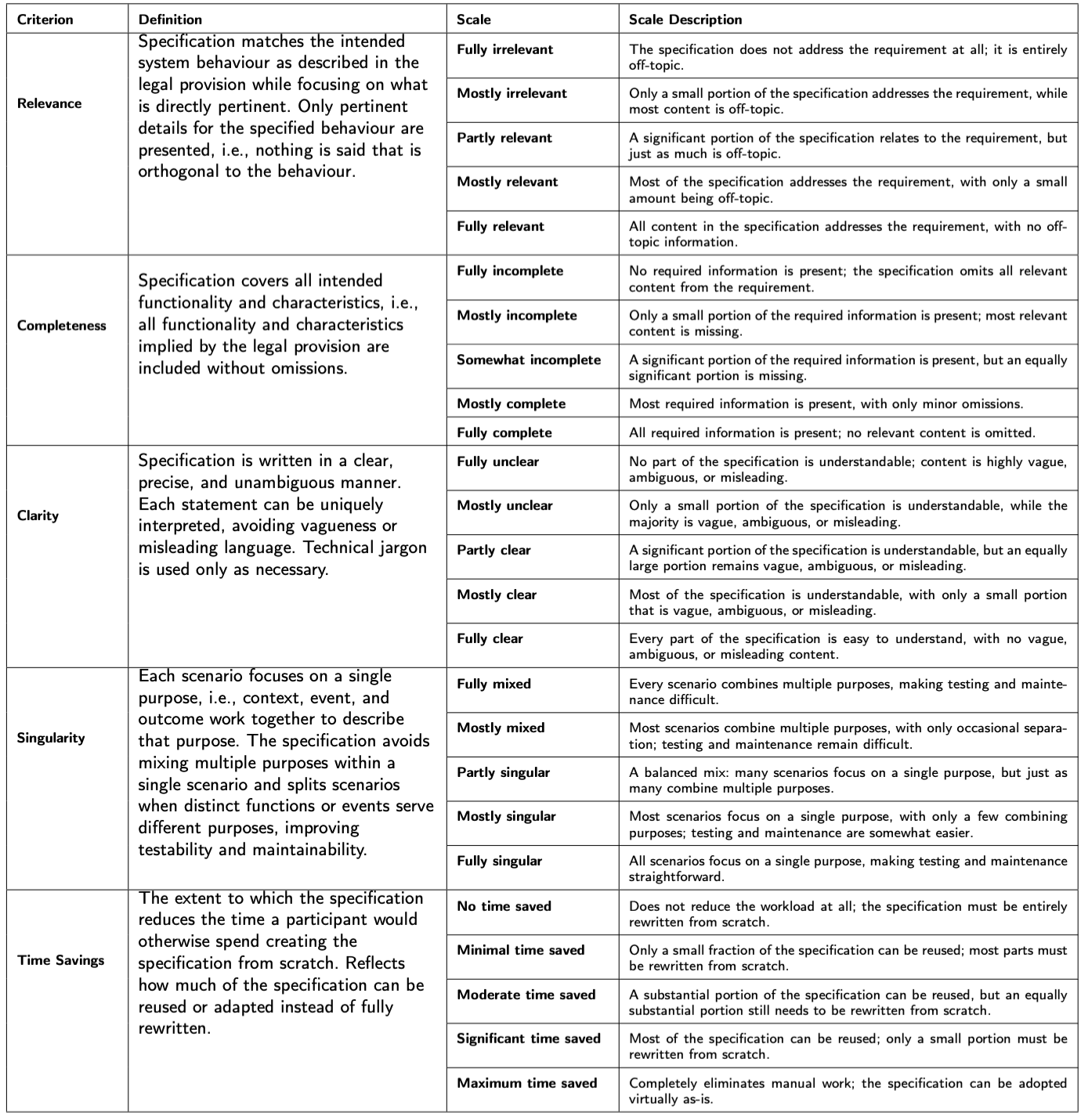}
\end{table*}

In the third step, participants evaluated each generated Gherkin specification against the defined quality criteria. 
Before performing their tasks, participants received:

\textit{Training materials: } 1)~A brief tutorial on BDD principles and Gherkin syntax, clarifying how Gherkin can be used for legal compliance; and 2)~Example Gherkin specifications that illustrate pitfalls and best practices (e.g., incomplete scenarios, and single purpose per scenario)~\cite{Data}.

\textit{Evaluation instruction: } 1)~A concise explanation of each quality criterion -- namely \emph{relevance}, \emph{completeness}, \emph{clarity}, \emph{singularity}, and \emph{time savings} -- with ordinal scale definitions to ensure consistent scoring (Table~\ref{tab:evaluation-criteria}), together with instructions to provide brief qualitative feedback when problems with the specification are identified. 
2)~Instructions to conduct a plausibility check, defined as a binary assessment of whether a specification could be realized in a real-world implementation. A specification is considered plausible if it could reasonably occur in practice without violating fundamental physical laws (e.g., conservation of mass) or containing logically inconsistent conditions, and if it aligns with common-sense expectations for the domain. Participants recorded their judgment as ``yes'' or ``no''.

\begin{figure*}
    \centering
\includegraphics[width=.6\linewidth]{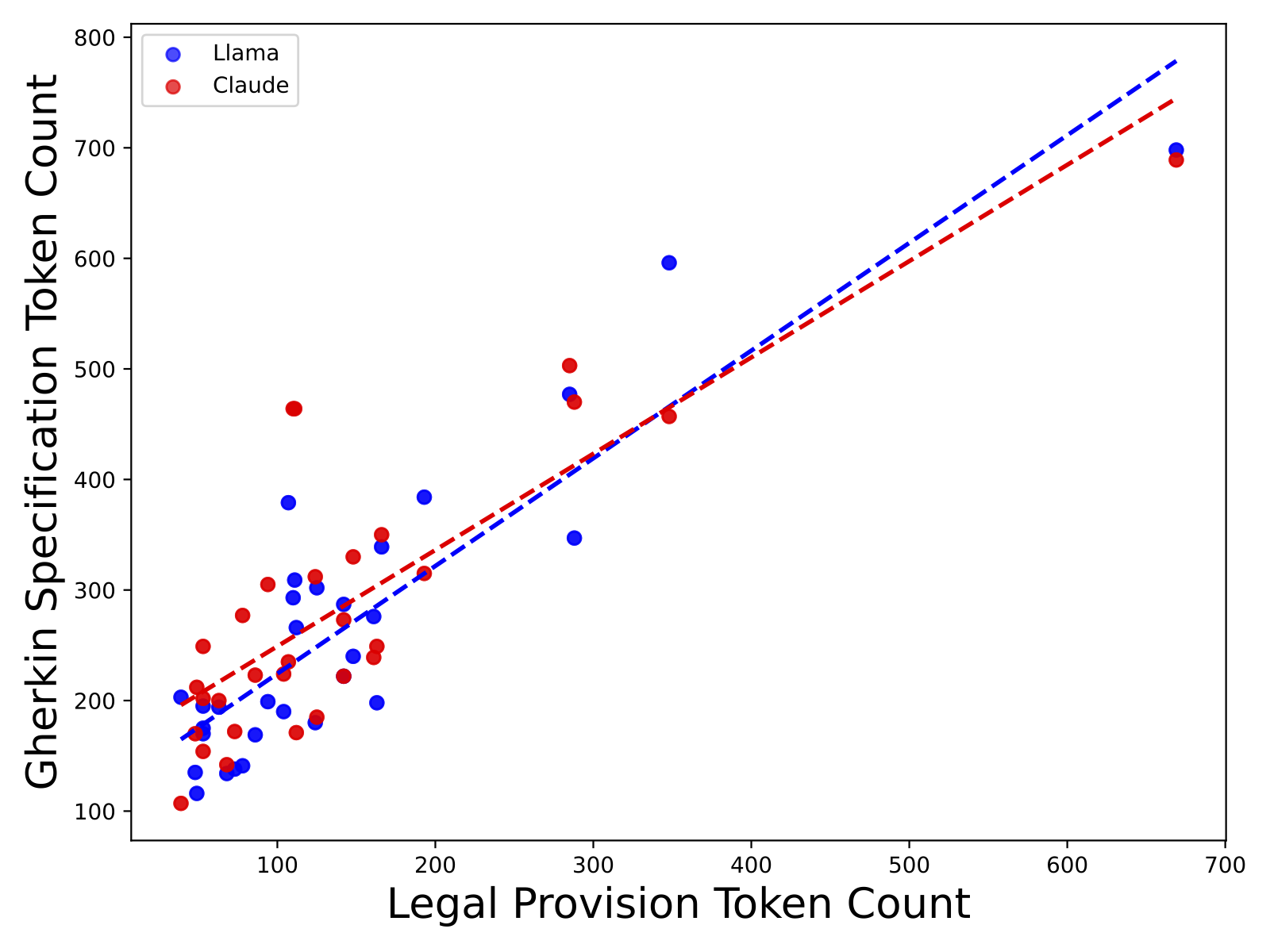}
    \caption{Scatter plot of legal-provision token counts versus corresponding Gherkin specification token counts for Llama and Claude, with fitted regression lines.}
    \label{fig3:scatterplot}
    
    \centering
    \includegraphics[width=\linewidth]{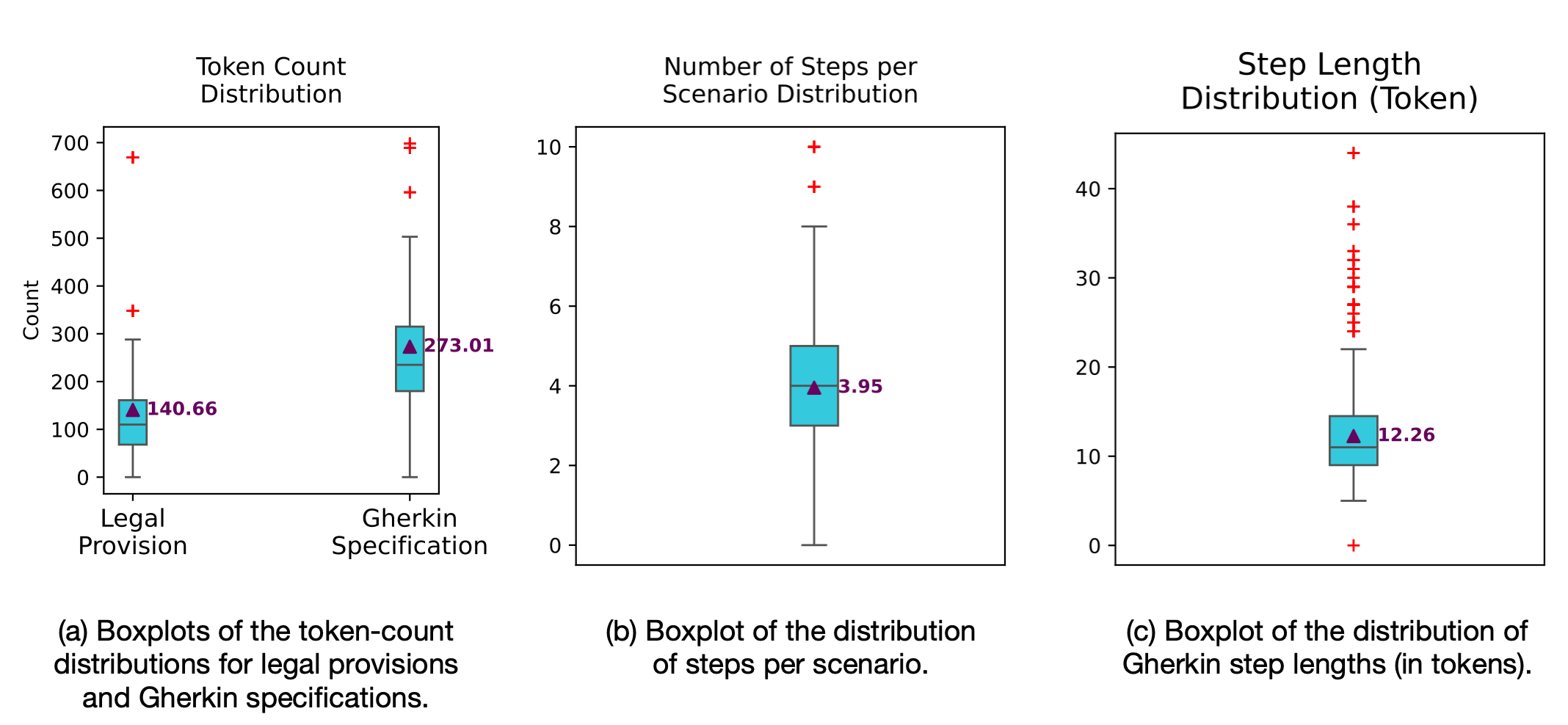}
\vspace*{-.5em}
\caption{Boxplots summarizing structural characteristics of legal provisions and Gherkin specifications.}\label{fig:boxplots}
\vspace*{-1em}
\end{figure*}

\section{Results} \label{sec:results}
A total of 60 Gherkin specifications were generated from food-safety legal provisions -- 30 by Claude and 30 by Llama. Specifications were evaluated by 10 participants, each of whom independently assessed 12 specifications (6 per LLM) across five quality criteria: \emph{relevance}, \emph{completeness}, \emph{clarity}, \emph{singularity}, and \emph{time savings} (Table~\ref{tab:evaluation-criteria}). \change{All five quality criteria are \emph{perceptual} rather than objective, as neither a gold standard nor time-tracking was used to assess Gherkin specifications in this study. Consequently, all ratings reflect participants' subjective perceptions.}{r3c7}{reviewer3Comment7}

The first four quality criteria originate from existing BDD scenario quality frameworks (Section~\ref{sec:BDDquality}) and are adapted for specification-level evaluation. The last criterion, \emph{time savings}, is proposed by us to evaluate participants' overall perception of how useful the generated specifications are. In addition to scale ratings, participants also provided qualitative feedback on the specifications.
No participant evaluated outputs from both LLMs for the same legal provision. Each specification was examined by two different participants (forming five distinct participant pairs), ensuring paired evaluations across all 60 specifications and resulting in 120 individual assessments.
All participants completed their tasks in full, and no data points were discarded.
The following subsections present the evaluation results, beginning with a descriptive analysis and subsequently addressing the research questions.

\subsection{Descriptive Analysis of Data}
We analyzed the input legal provisions and their corresponding Gherkin specifications by comparing token counts. Figure~\ref{fig3:scatterplot} presents a scatter plot of token counts for legal provisions vs. Gherkin specifications generated with both Llama and Claude. The plot also includes the fitted regression lines to highlight the relationship between the measures.
Figure~\ref{fig:boxplots}(a)~uses boxplots to compare the distributions of token counts in legal provisions and Gherkin specifications (Llama and Claude combined): legal provisions average $\approx 140.66$ tokens (median $=110$), whereas Gherkin specifications average $\approx 273.01$ tokens (median $=235$).
Llama-generated Gherkin specifications (median $\approx 212.5$, mean $\approx 265.1$ tokens) were shorter than those from Claude (median $= 244$, mean $\approx 285.5$ tokens). 

\change{We use the non-parametric Mann-Whitney $U$ test~\cite{Capon1991Elementary} to assess the statistical significance of differences between the distributions of two independent samples. 
}{r3c2}{reviewer3Comment2}
When comparing token counts between Llama- and Claude-generated Gherkin specifications, the observed difference was not statistically significant (Mann-Whitney $U$ test, $p = 0.18$).

We further examined the structure of the Gherkin specifications by analyzing the number of steps per scenario. As shown in Figure~\ref{fig:boxplots}(b), half of all scenarios contain between three and five steps (median $\approx 4$; mean $\approx 3.95$), with only a handful of high-end outliers extending to nine or ten steps.
Finally, to understand the granularity of each step, Figure~\ref{fig:boxplots}(c) presents a boxplot of step lengths measured in tokens. Typical steps span 9--15 tokens (mean $\approx 12.26$, median $\approx 11$). Occasional outliers extend \hbox{beyond these ranges.}

\subsection{Answers to RQs} \label{sec:Quantitative Evaluation}
Participants' ratings were provided on an ordinal scale. We report stacked bar plots for each evaluated quality criterion and compare participant- and LLM-level ratings using statistical significance tests.

\subsubsection{RQ1: How do LLMs perform when generating Gherkin specifications from \change{food-safety}{}{reviewer1Comment1} legal provisions?}
\label{subsubsec:RQ1}

\begin{sidewaysfigure*}
    \centering
    \includegraphics[width=24cm]{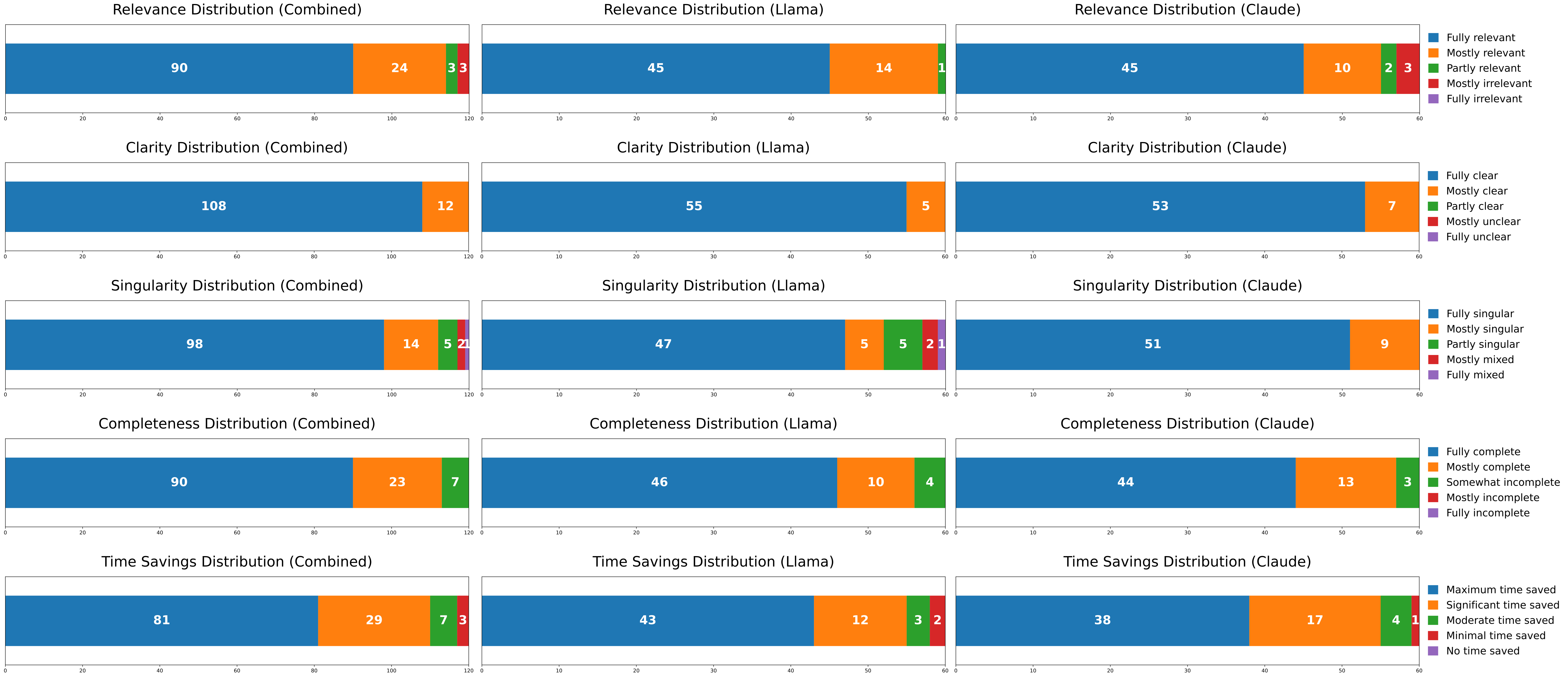}
    \caption{Stacked bar plots per criterion for all evaluations across both LLMs combined and stacked bar plots per criterion per LLM.}
    \label{fig3:stackedbarplotRQ1_RQ2}
\end{sidewaysfigure*}

In Figure~\ref{fig3:stackedbarplotRQ1_RQ2}, stacked bar plots display the distribution of participants' ratings across the evaluated quality criteria. The combined results are used to address RQ1. Each bar in the combined results represents 120 ratings for an individual criterion (6~tasks per model $\times$ 2 models $\times$ 10 participants). Overall, the plots indicate a strong consensus among participants’ ratings, with all criteria consistently receiving the highest possible median scores (``Fully clear'', ``Fully singular'', ``Completely relevant'', ``Fully singular'', and ``Maximum time saved''). Several observations \hbox{emerge from the plots.}

1)~\emph{Relevance: } 
The distribution of the ratings for \emph{relevance} indicates highly relevant Gherkin specifications to legal provisions. 
Of the 120 ratings, 75\% (90) are ``Fully relevant'' and 20\% (24) are ``Mostly relevant''. Only 2.5\% (3) fall into ``Partly relevant'' and 2.5\% (3) into ``Mostly irrelevant'',
indicating a small portion of off-topic information. No responses fall into the ``Fully irrelevant'' category.

2)~\emph{Clarity: }  
The distribution of the ratings for \emph{clarity} indicates highly clear Gherkin specifications. Of the 120 ratings, 90\% (108) are ``Fully clear'' and 10\% (12) are ``Mostly clear'', indicating a small fraction of specifications are ambiguous. No responses are ``Partly clear'', ``Mostly unclear'', or ``Fully unclear'', indicating the absence of substantial portions with confusion or difficulty in interpretation.

3)~\emph{Singularity: }
The distribution of the ratings for \emph{singularity} indicates single-purpose scenarios in most Gherkin specifications. Of the 120 ratings, 81.7\% (98) are ``Fully singular'', meaning each scenario targets exactly one purpose. 11.7\% (14) are ``Mostly singular'', indicating that while most scenarios maintained clarity of purpose, few scenarios contain multiple purposes. Only 4.2\% (5) are ``Partly singular'', 1.7\% (2) are ``Mostly mixed'', and 0.8\% (1)  ``Fully mixed'', indicating very few scenarios contain overlapping purposes.

4)~\emph{Completeness: } 
The distribution of the ratings for \emph{completeness} indicates that most Gherkin specifications capture the intended functions and characteristics implied by the legal provisions without omissions. Of the total 120 ratings, 75\% (90) are ``Fully complete'', 19.2\% (23) as ``Mostly Complete'', reflecting specifications with a small portion of relevant information being omitted. Only 5.8\% (7) are ``Somewhat incomplete'', and none are ``Mostly incomplete'' or ``Fully incomplete'', indicating most specifications include at least a substantial portion of the necessary information.
  
5)~\emph{Time savings: }
The distribution of the ratings for \emph{time savings} indicates substantial efficiency gains provided by Gherkin specifications. Of the 120 ratings, 67.5\% (81) are ``Maximum time saved'', reducing manual effort by eliminating extensive rewriting. 24.2\% (29) are ``Significant time saved'', showing small portions require rewriting. 5.8\% (7) are ``Moderate time saved'', showing substantial portions need editing. Only 2.5\% (3) are ``Minimal time saved'', indicating only small portions are reusable. No responses are ``No time saved'', thus there is some efficiency gain in all cases.

\begin{table*}
    \centering
    \caption{\textcolor{black}{Wilcoxon signed-rank} tests for each participant pair (two raters who scored the same 12 specifications) across the five quality criteria.  Cells report $p$-values, and \textcolor{black}{Benjamini-Hochberg-adjusted $p$-values ($p_{\textrm{BH}}$)}.}
    \small
    \includegraphics[width=0.75\linewidth]{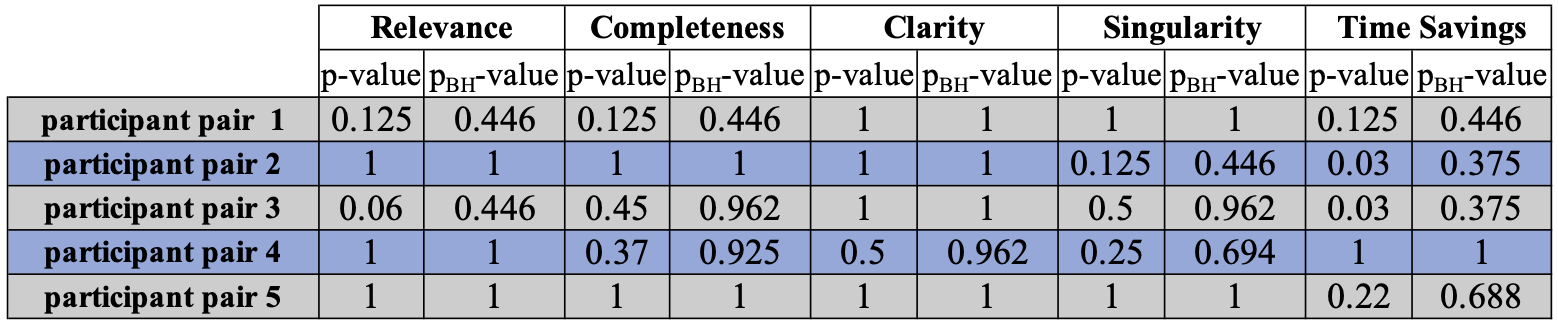}
    \label{tab:SignificanceRQ1}
\end{table*}

We recall that each participant pair corresponds to two participants who evaluate the same set of 12 Gherkin specifications. With ten participants, this yields five independent pairs. For each quality criterion, the two participants in a pair provide separate ratings, producing two matched sets of 12 ratings per pair for comparison. 
\textcolor{black}{We use the Wilcoxon signed-rank test~\cite{Capon1991Elementary} to compare paired ratings within participant pairs (matched by specification); this test is a suitable choice for paired ordinal data without assuming a normal distribution~\cite{Jamieson2004Likert,deWinter2010Likert}.}
\change{To control the false discovery rate (FDR) when performing multiple comparisons, we apply the Benjamini–Hochberg (BH) correction~\cite{Benjamini1995ControllingTF}.}{r3c1}{reviewer3Comment1}

\change{The $p$-values (and BH-adjusted $p$-values) in Table~\ref{tab:SignificanceRQ1} test the null hypothesis ($H_0$) that there is no systematic within-specification shift between the paired ratings (i.e., the paired changes are centred at zero). With two exceptions, none of the comparisons across the five  criteria -- \emph{relevance, completeness, clarity, singularity, \emph{and} time savings} -- shows a statistically significant difference. Specifically, participant pairs~2 and~3 yield nominally significant unadjusted $p$-values for \emph{time savings}; however, these effects do not survive BH correction. Thus, after BH adjustment, we find no statistically reliable differences for any quality criterion.}{r3c1-1}{reviewer3Comment1}

Regarding the plausibility of the specifications, we recorded four instances in which participants indicated that a generated specification was ``not plausible,'' with one case shared between two participants in a pair; we later present this case as Example~1 in Section~\ref{sec:discussion}. Therefore, there were only three unique specifications deemed implausible by participants. The assessment of plausibility was based on common sense and basic physical reasoning.

\begin{tcolorbox}[myframe]
\emph{The answer to {\bf RQ1} is:} Overall, the LLMs generated high-quality Gherkin specifications from food-safety legal provisions: across 120 human ratings, 95\% were mostly/fully relevant, 94\% mostly/fully complete, 100\% mostly/fully clear, 93\% mostly/fully singular, and 92\% reported significant/maximum time saved; only 3 unique specifications were judged implausible. There were no statistically reliable differences between participants.
\end{tcolorbox}

\subsubsection {RQ2: How do Llama and Claude compare when generating Gherkin specifications from \change{food-safety}{}{reviewer1Comment1} legal provisions?} 
\label{subsubsec:RQ2}

We use the per-LLM results in Figure~\ref{fig3:stackedbarplotRQ1_RQ2} to answer RQ2, presenting a direct model-to-model comparison across the evaluation criteria. Each bar in the per-LLM results reflects 60 ratings per model per criterion (6 tasks $\times$ 10 participants), highlighting differences in performance between Llama and Claude. Several observations emerge from the plots.

1)~\emph{Relevance: }
Both Llama and Claude perform well in generating relevant Gherkin specifications, with similar distributions of participant ratings. For Llama, 45 responses (75\%) are ``Fully relevant'', followed by 14 responses (23.3\%) as ``Mostly relevant'', and 1 response (1.7\%) as ``Partly relevant''. There are no ratings in the lowest two categories. 
Claude also achieves 45 ``Fully relevant'' ratings (75\%), with 10 (16.7\%) as ``Mostly relevant'', 2 (3.3\%) as ``Partly relevant'', and 3 (5\%) as ``Mostly irrelevant''. 
As with Llama, no responses are rated as ``Fully irrelevant''. While both models' outputs show high alignment with the legal provisions, Llama's output distribution is slightly more concentrated at the top, with fewer mid- or low-range responses than Claude. These results suggest that both LLMs are highly capable of producing relevant content to the legal provision, though Llama appears to have slightly better consistency in maintaining top-tier relevance.
 
2)~\emph{Clarity: }
Both Llama and Claude received high ratings for \emph{clarity}, with no responses falling into the bottom three categories. For Llama, 55 out of 60 ratings (91.7\%) are ``Fully clear'', indicating that nearly all specifications are perceived as unambiguous. The remaining 5 responses (8.3\%) are ``Mostly clear'', suggesting ambiguity in a few cases. Claude's ratings are similarly strong, with 53 responses (88.3\%) as ``Fully clear'' and 7 (11.7\%) as ``Mostly clear''. While both models are able to generate consistently clear output, Llama shows a slight advantage, with a greater proportion of top-tier ratings.

3)~\emph{Completeness: }
Both Llama and Claude received strong ratings for \emph{completeness}. For Llama, 46 responses (76.7\%) are ``Fully complete'', with an additional 10 (16.7\%) as ``Mostly complete''. Only four responses (6.7\%) are ``Somewhat incomplete''. Claude shows comparable performance with 44 responses (73.3\%) as ``Fully complete'', 13 (21.7\%) as ``Mostly complete'', and 3 (5\%) as ``Somewhat incomplete''. Neither model received ratings in the two bottom categories. Llama offers marginally more coverage of legal content as presented by ``Fully complete'' category (76.7\% vs. 73.3\%), although the distributions are close.

4)~\emph{Singularity: } 
Claude and Llama both perform well in maintaining scenario \emph{singularity}, though their rating distributions show differences. Claude received 51 ratings (85\%) as ``Fully singular'' and 9 (15\%) as ``Mostly singular''. No responses are rated as ``Partly singular'', ``Mostly mixed'', or ``Fully mixed''.
Llama received 47 ``Fully singular'' ratings (78.3\%) and 5 ``Mostly singular'' ratings (8.3\%). However, it also received more dispersed ratings with 5 (8.3\%) as ``Partly singular'', 2 (3.3\%) as ``Mostly mixed'', and 1 (1.7\%) as ``Fully mixed''. Overall, Claude shows more consistent adherence to the principle of scenario singularity.

5)~\emph{Time savings:}
 Claude and Llama are perceived to be effective at reducing the effort required to produce specifications from scratch, although their distribution differs slightly. For Llama, 43 responses (71.7\%) are ``Maximum time saved'',  suggesting that in most cases, the specifications could be reused with minimal editing. An additional 12 responses (20\%) are ``Significant time saved'', reflecting significant but not full reuse. Only 3 responses (5\%) are ``Moderate time saved'' and 2 (3.3\%) are ``Minimal time saved''.  
Claude showed slightly lower top-tier scores with 38 responses (63.3\%) as ``Maximum time saved'' and 17 (28.3\%) as ``Significant time saved''. Only 4 responses (6.7\%) are ``Moderate time saved'', and 1 response (1.7\%) as ``Minimal time saved''. Overall, both models are perceived to offer considerable time savings, with Llama perceived to be slightly ahead in producing outputs that participants could adopt almost as-is.

We apply the Mann-Whitney $U$ test to compare the outputs of Llama and Claude across our five quality criteria.
The null hypothesis ($H_0$) for each comparison states that there is no difference in the distribution of participant ratings between the two models for a given quality criterion. 
\change{The results in Table~\ref{tab:SignificanceRQ2} show that no statistically significant differences are observed between Llama and Claude on any quality criterion, under either the unadjusted or BH-adjusted $p$-values.}{r3c1-2}{reviewer3Comment1} 
These findings indicate that Llama and Claude perform comparably across all five evaluated quality criteria, with no evidence of either model outperforming the other in our study context.
\begin{table*}
    \centering
    \caption{Mann-Whitney $U$ tests comparing Llama and Claude across the five quality criteria. Cells report $p$-values, and \textcolor{black}{ Benjamini-Hochberg-adjusted $p$-values ($p_{\textrm{BH}}$)}.}
    \includegraphics[width=0.75\linewidth]{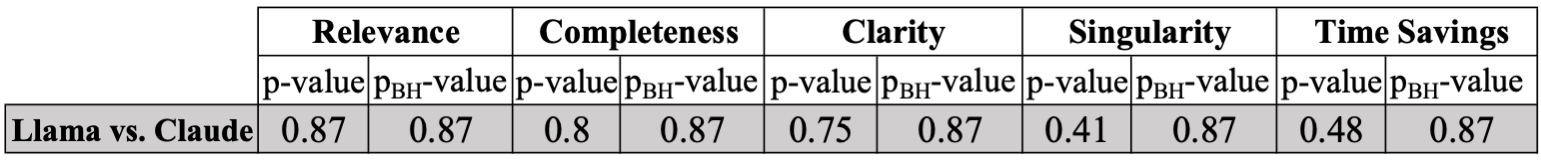}
    \label{tab:SignificanceRQ2}
\end{table*}

\begin{tcolorbox}[myframe]
\emph{The answer to {\bf RQ2} is:
Claude and Llama performed comparably; statistical testing found no significant differences between them on any quality criterion. Descriptively, Llama was slightly higher on clarity/completeness/time savings, while Claude was more consistently singular (fewer mixed-purpose scenarios).}
\end{tcolorbox}

\section{Analysis of Qualitative Feedback} \label{sec:discussion}
To complement the quantitative results presented in Section~\ref{sec:results}, we now turn to a \emph{qualitative} analysis of the textual feedback provided by participants in order to gain insights into the types of issues that arise and the reasoning behind their judgments. In this section, we first summarize descriptive statistics on the comments received, then present the key themes that emerged from the analysis, and finally illustrate the most salient themes with representative examples.

\subsection{Summary Statistics}
Across 120 assessments, participants left a total of 105 comments. Of these, 64 (61\%)  pointed out potential issues in the generated specifications, while 41 (39\%) were approvals or alternative suggestions, even though all criteria had been assigned the highest score. Thus, nearly two thirds of the feedback identified specification issues. 

Breaking statistics down per participant, each participant contributed between 5 and 12 comments each (mean $= 10.5$, SD $= 2.6$), with seven participants providing the maximum of 12 comments. When filtering out non-issue comments (approvals or suggestions), the number of issue-specific comments per participant ranged from 3 to 11 (mean $= 6.4$, SD $= 2.6$), with one participant providing the maximum of 11 unique issue-specific comments. The number of approval comments ranged from 1 to 8 per participant (mean $= 4.1$, SD $= 2.7$), with two participants providing the maximum of 8 approval comments. Every one of the 10 participants provided both issue-specific comments and approval comments.

Aggregated by quality criterion, we obtained 39 comments related to \emph{time savings}, 30 to \emph{relevance}, 30 to \emph{completeness}, 22 to \emph{singularity}, and 12 to \emph{clarity}.
\emph{Clarity} received input from 7 unique participants, while the other criteria received feedback from 9 unique participants each.

\subsection{Recurring Themes in Participant Comments}
We examined the textual feedback from participants through a close reading of all comments and the identification of recurring themes, which revealed common sources of concern such as irrelevant details, omissions, and ambiguities. The main themes that emerged pertained to three quality criteria: \emph{relevance}, \emph{completeness}, and \emph{singularity}.

For \emph{relevance}, participants noted issues such as off-topic or hallucinated content not grounded in the legal text, unnecessary repetition of information, and misinterpretations where the model generated inaccurate statements.  
Overall, our analysis shows that generated specifications sometimes contained unnecessary assumptions about system behaviours. Observed patterns further indicated that scenario creation was at times influenced by constraints and practices not present in the source, which affected the perceived \emph{relevance} of the specifications. 

For \emph{completeness}, the most common concerns were omissions of entire requirements, omissions of specific clauses, or omissions of details.  
Overall, our analysis shows that generated specifications on occasion omitted necessary details and phrases. In some cases, functions and characteristics supported by the legal provision were not clearly reflected in the specification. These omissions contributed to misinterpretations and increased participants' effort in rewriting, which negatively affected the \emph{completeness}. 

Finally, for \emph{singularity}, when legal provisions were lengthy or covered multiple obligations (e.g., size, mass, or temperature measurements), participants identified situations where scenarios were not split, making it harder to preserve a single purpose. Conversely, some comments pointed to unnecessary scenario splitting, where combined scenarios would have been better. 
Overall, our analysis shows that when legal provisions were long and complex, generated specifications sometimes contained multiple purposes within the same scenario, which reduced \emph{singularity} and also complicated \emph{clarity}. In other cases, excessive splitting created scenarios that were fragmented and less comprehensible. Both tendencies introduced ambiguity about how the specification related to the legal text, increasing the risk of misinterpretation.

\begin{figure}
    \centering
    \includegraphics[width=\linewidth]{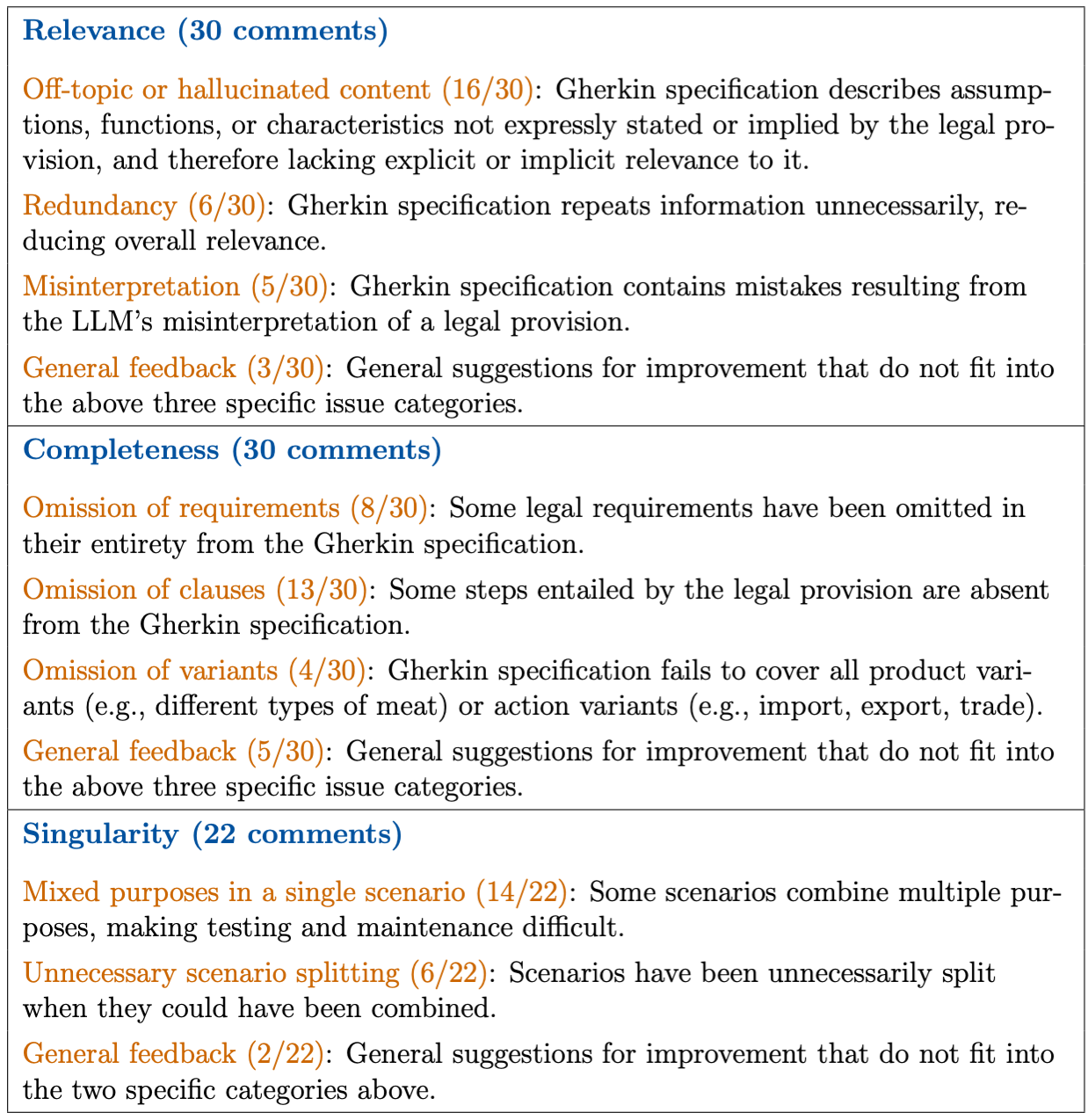}
    \caption{Issue themes observed in LLM-generated Gherkin specifications, grouped by quality criteria. Numbers in parentheses indicate the counts of participant comments. An individual comment could relate to multiple quality criteria if it addressed more than one type of issue simultaneously.}
    \label{tab:relevance_completeness_singularity}
\end{figure}

Figure~\ref{tab:relevance_completeness_singularity} presents a  breakdown of participant feedback on \emph{relevance}, \emph{completeness}, and \emph{singularity}, along with the frequency of each identified theme. For each quality criterion, the figure also includes a ``general feedback'' category for comments that did not address a specific deficiency. In the remainder of this section, we illustrate some of the themes from Figure~\ref{tab:relevance_completeness_singularity} using three legal provisions, their corresponding Gherkin specifications, and selected participant feedback.

The first example, Example~1, focuses on the issues observed in relation to \emph{relevance}:

\begin{tcolorbox}[colback=gray!5,colframe=black!50!white,title=Example 1 -- Legal Provision]
(5) In the case of an inspection of a consumer prepackaged food that consists of a liquid, the net quantity of the food must be determined on the basis of the assumption that the liquid is at a temperature of \SI{20}{\celsius}.
\end{tcolorbox}

\begin{figure*}
    \centering
    \includegraphics[width=\linewidth]{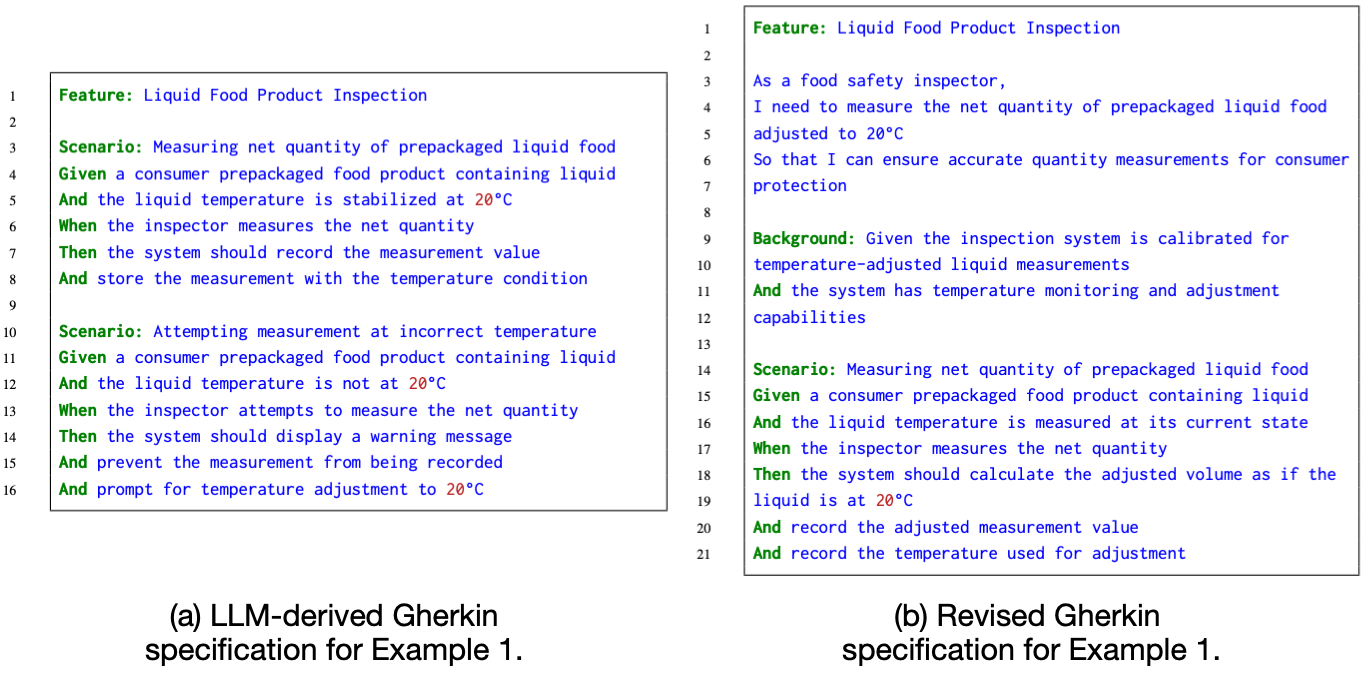}
    \caption{LLM-derived and revised Gherkin specifications for Example 1.}
    \label{fig:example1}
\end{figure*}

Figure~\ref{fig:example1} (a)~shows an LLM-derived Gherkin specification for Example 1. The \emph{relevance}-related feedback from the two participants who examined this specification is as follows.

\begin{tcolorbox}[colback=gray!5,colframe=black!50!white,title=Feedback (Example~1) -- Participant 1]
I think that the second scenario is irrelevant to the requirement content as it relies on assumptions about the system (such as assuming it will display a warning). However, it's unclear whether the system even includes a visual i/o component.
\end{tcolorbox}


\begin{tcolorbox}[colback=gray!5,colframe=black!50!white,title=Feedback (Example~1) -- Participant 2]
I find the content of the second scenario to be irrelevant and include hallucinations as the legal requirements does [sic] not talk about displaying a warning. Also, the assumption that all liquids with prepackaged food can be held at exactly 20 degrees is far-fetched. The concept of matching the volume in current temperature to what it should be at 20 degrees is much more feasible.
\end{tcolorbox}

Here, both participants critique the second scenario. One argues that requiring the system to display a warning message and prevent measurement goes beyond what the provision could reasonably entail, and should therefore be discarded to improve \emph{relevance}. The second participant observes that the specification should reflect what is plausible in real-world practice, noting that physically bringing a liquid to a temperature of \SI{20}{\celsius} merely for the purpose of measuring its volume is impractical. Instead, they conclude that the most reasonable interpretation is that the volume should be (mathematically) adjusted as if the liquid were at \SI{20}{\celsius}. Based on the feedback received, a revised Gherkin specification, which more accurately reflects realistic measurement practices, is presented in Figure~\ref{fig:example1} (b).

\begin{figure*}
    \centering
    \includegraphics[width=\linewidth]{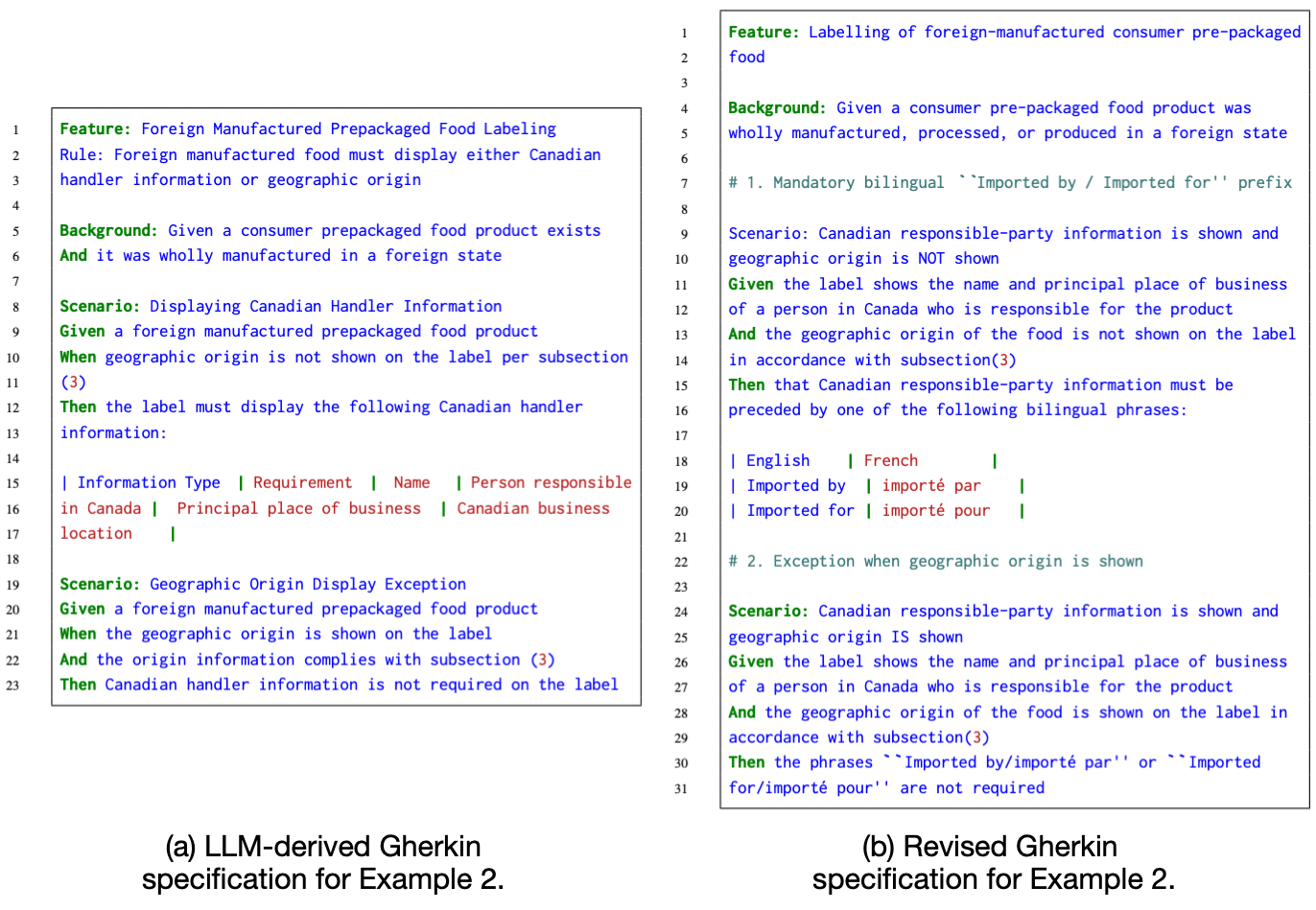}
    \caption{LLM-derived and revised Gherkin specifications for Example 2.}
    \label{fig:example2}
    \vspace*{-1em}
\end{figure*}

Our second example, Example~2, concerns issues observed in relation to \emph{completeness}. The feedback from participants on the LLM-derived Gherkin specification for Example~2, shown in Figure~\ref{fig:example2}(a), indicates that several omissions are present.

\begin{tcolorbox}[colback=gray!5,colframe=black!50!white,title=Example 2 -- Legal Provision]
223 (1) If a consumer prepackaged food was wholly manufactured, processed or produced in a foreign state and the name and principal place of business of the person in Canada for whom it was manufactured, processed or produced or the person by whom it was stored, packaged or labelled in Canada is shown on its label, that information must be preceded by the expressions ``Imported by'' and ``importé par'' or ``Imported for'' and ``importé pour'', as the case may be, unless the geographic origin of the consumer prepackaged food is shown on the label in accordance with subsection (3).
\end{tcolorbox}

\begin{tcolorbox}[colback=gray!5,colframe=black!50!white,title=Feedback (Example~2) -- Participant 1]
The ``import'' expression listed in the requirement did not appear to have a scenario associated to it. I've provided my version of this new scenario.
\end{tcolorbox}

\begin{tcolorbox}[colback=gray!5,colframe=black!50!white,title=Feedback (Example~2) -- Participant 2]
I find that the generated specification misses what the requirement was discussing and interprets it incorrectly. Based on my understanding the requirement is discussing when ``imported for'' or ``imported by'' should be a ``label data''. But the generated specification has a scenario that does not mention this ``label data'' and focuses on if the name of the person should be a ``label data''. The specification would be more complete if it contained all terms ``manufactured'', ``processed'' or ``produced''. I didn't find ambiguity other than what was mentioned. 
\end{tcolorbox}

Taken together, the feedback suggests that the LLM-derived specification is missing some important legal terms and fails to capture the intended labelling requirements. In particular, the specification does not adequately reflect the mandatory bilingual expressions or the conditions under which they apply. To address the incompleteness, a revised version that more faithfully represents the provision is presented in Figure~\ref{fig:example2}(b).

Our third and final example illustrates issues related to \emph{singularity}, particularly the tendency of LLMs to generate multi-purpose scenarios. Due to their length, the underlying legal provision, the LLM-derived Gherkin specification, and the revised Gherkin specification are provided in Appendix~\ref{sec:appendix-example3}. The feedback on \emph{singularity} from the two participants who reviewed the LLM-derived Gherkin specification (Figure~\ref{fig:example3original}) is as follows:

\begin{tcolorbox}[colback=gray!5,colframe=black!50!white,title=Feedback (Example~3) -- Participant 1]
Response has stacked all of the sub-requirements into two specifications, per equine and per bird. Many details have been omitted by the Gherkin for the table. The answer is totally wrong. A Detailed version is provided.
\end{tcolorbox}

\begin{tcolorbox}[colback=gray!5,colframe=black!50!white,title=Feedback (Example~3) -- Participant 2]
I can add similar elements as the background and write different parts in each scenario. The main problem I had with this one is that (h) is only talking about equine (not bird). But the second scenario has included subitems of (h) for a bird, which is incorrect.
\end{tcolorbox}

The main problem with this example is that the LLM-derived Gherkin specification aggregates all sub-requirements into two scenarios, each with a large ``Then'' step including an information table, rather than producing one scenario per individual obligation. For example, sub-clauses (e)(i), (e)(ii), and (e)(iii) concerning birds should each appear as separate scenario when they represent distinct checks. 

In addition to this primary problem with \emph{singularity}, the LLM-derived specification also entirely omits several nested items. For instance, with respect to birds, it fails to include recovery dates for diseases, extra-label drug prescription attestations, and other details specified under clause (g) sub-items. These missing elements fall under the \emph{completeness} criterion. Since similar omissions were already illustrated in Example~2, we do not elaborate further here.

As one possible revision based on the participants' feedback, 
Figures~\ref{fig:example3modified}(a) and \ref{fig:example3modified}(b) 
in Appendix~\ref{sec:appendix-example3} present a case where the 
requirements for equines and birds are separated into two distinct 
feature files, each containing single-purpose scenarios.

\section{Discussion} \label{sec:practical}
\textcolor{black}{In this section, we discuss the criticality of LLM generation errors for the task investigated in this article as well as the broader implications of our findings.}
\change{\subsection{LLM Failure Risks}\label{subsec:failure}
The issues observed in our qualitative analysis of LLM-generated Gherkin specifications, namely missing clauses, hallucinated content, and conflated intents, carry different levels of risk. In particular, in safety-critical contexts such as food handling, omissions appear to be the most consequential risk: unless caught by humans, the LLM overlooking a regulatory constraint can lead to potentially unsafe outcomes that nonetheless pass testing, only to fail under real-world regulatory scrutiny or operational conditions. In contrast, hallucinated content, although still serious, is easier for human reviewers to identify, given the expectation that LLMs are prone to generating spurious content. Multi-purpose scenarios, while undesirable, are secondary to omissions and hallucinations, as they are likely to be naturally identified and resolved during downstream refinement activities. The \emph{main takeaway} here is that automated derivation of Gherkin specifications from laws and regulations -- like most other LLM-based automation -- cannot be assumed to be fully accurate, with omissions and hallucinations remaining notable risks. Educating users about these risks is a key mitigation measure; for the same reason, human-in-the-loop verification is especially important in safety-critical domains where non-compliance could lead to harm to individuals, the environment, or society at large.}{r2c3}{reviewer2Comment3}\change{}{}{reviewer1Comment4}

\change{\subsection{Implications for Research} \label{subsec:research implications}
Our study provides empirical evidence that existing LLMs can support the transition from laws and regulations to behavioural specifications in a developer-centric format. These results broaden the feasible scope of automation for legal requirements and suggest that legal texts can be used not only for verbatim extraction or classification, but also as direct sources for structured software engineering artifacts. Methodologically, our study provides a reusable blueprint for evaluating LLM-generated artifacts using human judgment. At the same time, our qualitative findings highlight important limitations that should be addressed in future research. The issues discussed in Section~\ref{subsec:failure} point to the need for techniques that (i) ensure faithfulness to the source provisions, (ii) support systematic decomposition of long or cross-referenced clauses into reviewable units, and (iii) make the transformation process more explainable. This motivates research on prompt and glossary design as well as human-AI review protocols, including critique-refine loops, tailored to compliance-driven requirements analysis.}{r3c3}{reviewer3Comment3}

\change{\subsection{Implications for Practice} \label{subsec:practiceimplications}
Overall, and while strictly scoped to our study domain (as is typical for case-study research), our results suggest that LLMs at their current level of maturity can serve as effective co-authors of behavioural specifications in regulated domains, especially as a \emph{first-draft} mechanism. High ratings across the quality criteria, together with strong perceived time savings, indicate that teams should be able to start from LLM-generated specifications rather than writing scenarios from scratch. In practice, this shifts effort towards review and traceability management, rather than initial authoring.
However, practitioners should assume that LLM outputs are \emph{draft} artifacts rather than authoritative statements of obligations. Several participant comments flagged issues (Section~\ref{sec:discussion}). Consequently, effective adoption should be framed as a human-in-the-loop workflow: (i)~scope and, where necessary, decompose provisions into smaller fragments; (ii)~build a domain glossary to establish terminology; (iii)~generate candidate specifications using controlled prompts; and (iv)~review outputs against a checklist reflecting relevance, completeness, clarity, and single-purpose scenario structure, complemented by a plausibility check for unrealistic behaviours.
Teams should keep legal inputs, prompts, and generated outputs under configuration management to support auditability and reproducibility, and apply extra scrutiny to lengthy or cross-referenced provisions, where the risk of omissions is higher. When outputs contain implausible behaviour or hallucinations, teams should treat this as a signal to rescope the input or revise the prompting/glossary strategy, not merely as a one-off defect to patch.}{r3c3-1}{reviewer3Comment3}

\section{Validity Considerations}\label{sec:threats}
Below, we discuss validity considerations for our study, focusing on internal, external, construct, and statistical conclusion validity, and describing the measures we took to mitigate threats.

\emph{Internal validity.}
An important threat to internal validity is that participants may have rated the outputs differently simply because they knew they were evaluating LLM-generated specifications. We reduced this risk by keeping LLM identities hidden, providing standardized training with clear scoring rubrics and examples, and using two independent raters per specification. 

\change{A second threat is that participants may not have fully considered downstream consequences when judging the specifications. Participants understood that food handling is a safety-critical domain, but we did not overstress potential real-world harms from errors, since doing so could have shifted participants away from their natural decision-making. This creates a trade-off: ratings may reflect how well the specifications match the legal provisions and the predefined quality criteria more than how they would be scrutinized in a consequence-focused compliance setting. To keep evaluations grounded in reality, we included a plausibility check that asked whether the specifications could reasonably materialize in a real-world implementation. We also required brief qualitative feedback when participants identified problems.}{r1c6}{reviewer1Comment6} 

A third threat is order-related carryover effects. While each participant rated any given  provision only once (i.e., never comparing outputs from both models for the same provision), earlier tasks could still influence how they used the rating scale on later, unrelated provisions. To address this, we randomized task order for each participant and balanced model assignments across sequences so that exposure to the two LLMs was evenly distributed rather than clustered.

\change{A fourth threat is participant expertise. We deliberately recruited participants with a software engineering background rather than legal or food-safety expertise, because evaluating Gherkin specifications requires software engineering competence (e.g., familiarity with BDD conventions, testability, and scenario structure) that domain experts do not necessarily have. Nonetheless, residual risks remain: although participants were not asked to adjudicate legal intent or enforcement, assessing whether a scenario is faithful to what is stated in a legal provision still requires interpretation, thus creating a risk of misinterpretation. We mitigated this risk by having the author team (who have worked on food safety and the legal framework in scope for over four years and have interacted with food-safety experts) available during the study to provide interpretive assistance with the provisions. Yet, legal or food-safety professionals may identify nuances (e.g., scope, implicit exceptions, term definitions) that software engineers could miss; outcomes may therefore differ with mixed teams, making replication with mixed participants a natural next step.}{r1c3}{reviewer1Comment3}

\emph{External validity.}
Our findings are bounded by our study context: food-safety regulations from North American jurisdictions, 10 student participants from a single university, and two specific LLMs used with a fixed prompt template. To avoid overgeneralization, we limit our claims to this setting; within this scope, of course, we ensured that all participants had balanced exposure to both LLMs. Although all participants came from the same institution, they ranged from senior undergraduates to final-year PhD candidates, providing a diversity of experience levels and reducing the likelihood that our results reflect a single skill level.

\change{In our prompt design, we assigned a consistent ``senior requirements engineering expert'' persona in the system message across all generations to reduce prompt-induced variability and to reflect a plausible software engineering stakeholder who would consume and refine behavioural specifications. Nevertheless, persona framing can steer an LLM's emphasis and style (e.g., prioritizing testability, completeness, or compliance interpretation), which may in turn influence both the structure and the content of the resulting  specifications. Alternative roles such as a quality engineer, compliance officer, legal counsel, or business analyst could lead to different trade-offs and potentially different outcomes~\cite{wu2023large,kong2024better}.}{r3c5-1}{reviewer3Comment5}

\change{The provisions in our study are primarily made up of operational requirements (e.g., measurable constraints and concrete obligations). Laws and regulations stated at higher levels of abstraction -- for example, those articulating broad principles, rights, or discretionary standards -- may require substantial contextual interpretation before they can be operationalized. Such provisions can have intricate cross-references, exceptions, and open terms that do not map as naturally to a \texttt{Given-When-Then} pattern.}{r2c4}{reviewer2Comment4}

\change{A further limitation concerns the replicability of our results. Replications with LLMs can be challenging because outputs may vary across runs and over time due to factors such as stochastic generation, provider-side updates, and model-version changes, even when prompts are held constant. Recent work has begun to propose reporting and packaging guidelines to mitigate some of these issues~\cite{baltes2025Guidelines}. While such guidelines are still evolving and not yet widely established, we documented our experimental configuration as precisely as possible and provide a replication package to support re-analysis and partial regeneration.}{r1c11}{reviewer1Comment11}

\emph{Construct validity.}
We evaluated five quality criteria on ordinal scales, plus a binary plausibility check. \emph{Time savings} is self-assessed rather than time-tracked, and the plausibility check is coarse (Yes/No). We constrained drift by supplying concise definitions, clear scale descriptions, and  examples during training, so participants applied shared interpretations when scoring each criterion. Using established BDD-inspired criteria (\emph{relevance}, \emph{completeness}, \emph{clarity}, \emph{singularity}) further grounded the constructs in prior literature, and we applied the same criteria uniformly across all tasks.

\emph{Statistical conclusion validity.}
The sample size (10 participants; 60 specifications; 120 ratings) limits power and introduces some clustering (multiple ratings per participant and provision pairings). To maintain appropriate inference under these constraints, we used non-parametric tests suited to ordinal data, reported p-values, \textcolor{black}{adjusted for multiple comparisons}, and applied the same analysis workflow across different quality criteria and comparisons.

\section{Related Work}\label{sec:related}
We review related work on LLM-assisted RE and on BDD quality assurance.

\subsection{LLMs for RE}
LLMs are increasingly used for requirements automation across three themes: \emph{analysis and assurance}, \emph{generation and transformation}, and \emph{legal reasoning and compliance}.

\subsubsection{Analysis and Assurance}
LLMs automate tasks such as ambiguity detection~\cite{arora2310advancing}, classification~\cite{Hey2020NoRBERT,Chatterjee2021Pipeline,Deshpande2021BERT,marques2024using}, obligation extraction~\cite{Sainani2020Extracting,jain2023transformer}, smell and dependency detection~\cite{Habib2021Detecting,Wang2022Detecting}, and completeness checking~\cite{luitel2024improving}. More recently, in-context reasoning has been used for quality assurance~\cite{Fazelnia2024Lessons} and satisfiability checks~\cite{Santons2024InContext}.

\subsubsection{Generation and Transformation} \label{sec:generation}
Rahman et~al.~\cite{rahman2024automated} convert requirements into user stories; Ronanki et~al.~\cite{ronanki2023investigating} and G\"{o}rner et~al.~\cite{Gorer23Generating} generate interview scripts. Nayak et~al.~\cite{Nayak2022Req2Spec}, Spoletini and Ferrari~\cite{spoletini2024return}, and Ferrari and Spoletini~\cite{ferrari2025formal} map natural language into formal or domain-specific languages (DSLs), while Wei et~al.~\cite{wei2024requirements} generate code stubs and test scaffolds.  

Lutze et~al.~\cite{lutze2024generating} enrich product catalogues and tailor specifications for ambient assisted living devices, highlighting prompt-engineering challenges. Xie et al.~\cite{xie2023impact} show few-shot learning improves specification fidelity, and Ronanki et al.~\cite{ronanki2023investigating} find LLM-generated requirements often surpass human ones. Ferreira et al.~\cite{Ferreira2025Acceptance} generate Gherkin scenarios from user stories, produce executable tests, and assess them through user feedback.  

Wang et al.~\cite{wang2025multi} derive acceptance criteria from multi-modal data and evaluate them by relevance, completeness, understandability, and coverage. Shi et al.~\cite{shi2024using} refine Gherkin specifications using stakeholder feedback, showing that while LLMs can produce detailed scenarios, human validation remains necessary. Fernandes de Sousa~\cite{sousa2025experimento} rewrites informal tests into Gherkin scenarios, reinforcing industrial interest in AI-assisted BDD.

These studies transform user stories, DSLs, code, and even media such as videos, but none derive Gherkin specifications from legal provisions.

\change{Beyond LLM-based generation, empirical requirements engineering has a long history of human-centred studies on requirements quality. Mund et al.~\cite{mund2015does} combine surveys and controlled experiments to examine when requirements quality affects downstream development tasks. Winter et al.~\cite{winter2020quantifiers} use a web-based experiment to study how quantifier formulations influence requirements readability and error rates. More recently, Frattini et al.~\cite{frattini2025applying} report a controlled experiment quantifying the effects of passive voice and ambiguous pronouns on domain-model derivation. Methodologically, our quasi-experiment follows this line of work through human-based analysis of requirements-related artifacts, but differs in two key respects: first, we examine LLM-generated Gherkin specifications derived from legal provisions, thus introducing a legal-compliance perspective; second, we focus on BDD-inspired quality criteria rather than modelling accuracy or reading performance, thus more specifically catering to agile and BDD-oriented development practices.}{r3c6}{reviewer3Comment6}

\subsubsection{Legal Reasoning and Compliance} \label{sec:compliance}
%
\change{Dahl et al.~\cite{dahl2024large} assess LLMs on specific factual questions about federal case law, reporting hallucination rates between 58\% and 88\%. Similarly, Magesh et al.~\cite{magesh2025hallucination} evaluate commercial retrieval-augmented legal research tools, finding that 17\% to 33\% of responses contain incorrect legal statements or unsupported citations, risks that have led to real-world sanctions for practitioners~\cite{Guardian2025LawyerAI}.
Our work differs fundamentally by focusing on a structurally distinct, syntax-based translation, mapping legal provisions to Gherkin specifications, rather than the open-world knowledge retrieval and legal reasoning that often leads to these errors.}{r1c5}{reviewer1Comment5} \change{}{}{reviewer1Comment11}

Singhal et al.~\cite{singhal2025legal} prompt LLMs to generate Python code instantiating a legal metamodel from statutory texts. In contrast, we generate and evaluate Gherkin specifications from legal texts.  
Kesari et al.~\cite{kesari2025legal} operationalize privacy laws by generating use cases, classifying compliance, and repairing violations. Unlike our work, theirs remains centred on structured use cases.  
Karpurapu et al.~\cite{karpurapu2024comprehensive} generate BDD acceptance tests, showing few-shot prompting improves quality. While they optimize agile BDD prompting, we transform legal provisions into Gherkin.  
Gokhan et al.~\cite{gokhan2024regnlp} present a regulatory question-answering system with completeness metrics; Fuchs et al.~\cite{fuchs2024using} translate regulations into logic; Abualhaija et al.~\cite{abualhaijallm} use RAG to extract GDPR obligations.  

These studies underscore LLM versatility in compliance analysis, yet none derive or evaluate behavioural specifications from legal texts through human-subject experiments. Our work helps fill this gap via a \hbox{quasi-experimental study}.

\subsection{BDD Specification Quality Assurance}\label{sec:BDDquality}
Ensuring high-quality user stories and BDD scenarios is difficult due to redundancy and ambiguity, leading to frameworks and guidelines for writing and evaluation.  

Heck et al.~\cite{heck2014quality} propose a framework for agile requirements with criteria of completeness, uniformity, and conformance, supported by a checklist covering clarity, rationale, identifiers, atomicity, and principles such as SMART~\cite{doran1981smart} and INVEST~\cite{wake2003invest}. Lucassen et al.~\cite{lucassen2015forging,lucassen2016improving} propose the Quality User Story framework, operationalized via an NLP-based quality-assurance tool.  

For BDD, Oliveira et al.~\cite{oliveira2017empirical,oliveira2019evaluate} propose a checklist including uniqueness, integrity, essentiality, singularity, completeness, clarity, focus, and ubiquity; Wautelet et al.~\cite{wautelet2023investigating} emphasize uniqueness, essentiality, integrity, and singularity. Mapping studies~\cite{binamungu2023behaviour} and suite-level principles by Binamungu et al.~\cite{binamungu2020characterising} ensure BDD suites remain coherent, adaptable, and understandable, though our study does not target suites. In our study, we adapted the criteria of relevance (aligned with essentiality~\cite{oliveira2019evaluate}), completeness, clarity, and singularity, and added time savings as a new criterion (Table~\ref{tab:evaluation-criteria}).

\section{Conclusion} \label{sec:conclusion}
\textcolor{black}{This article reported a human-subject study examining whether two LLM families (Claude and Llama) can translate food-safety regulations into Gherkin specifications that remain faithful to the source text. Across 120 assessments by 10 participants, the generated specifications were rated strongly on \emph{relevance}, \emph{clarity}, \emph{completeness}, \emph{singularity}, and perceived \emph{time savings}, suggesting that LLMs can serve as effective ``first-draft'' co-authors for legal requirements in our study context. Claude and Llama performed well overall, but qualitative feedback revealed important failure patterns -- omitted clauses, hallucinated details, and multi-intent scenarios -- suggesting that the real value lies in accelerating specification drafting with disciplined human review, not hands-off automation. Future work will explore transfer to other legal domains (e.g., privacy and data protection), ground generation in organizational artifacts (documents, process models, logs) to reduce omissions, make the workflow more interactive through human-in-the-loop feedback, and evaluate practical usefulness with practitioners.}

\section{Data Availability}\label{sec:package}
Our replication package is available on GitHub~\cite{Replication}, including the dataset~\cite{Data}, statistical analysis results~\cite{EvaluationResults}, and code for additional algorithms and evaluation scripts~\cite{Code}. Upon acceptance, the package will also be deposited on Zenodo for long-term preservation and accessibility.

\bibliographystyle{cas-model2-names}

\bibliography{bibfile}

\onecolumn

\appendix

\section{Supplementary Material for Section~\ref{sec:discussion}} \label{sec:appendix-example3}

\begin{figure}[h]
    \centering
    \includegraphics[width=.8\linewidth]{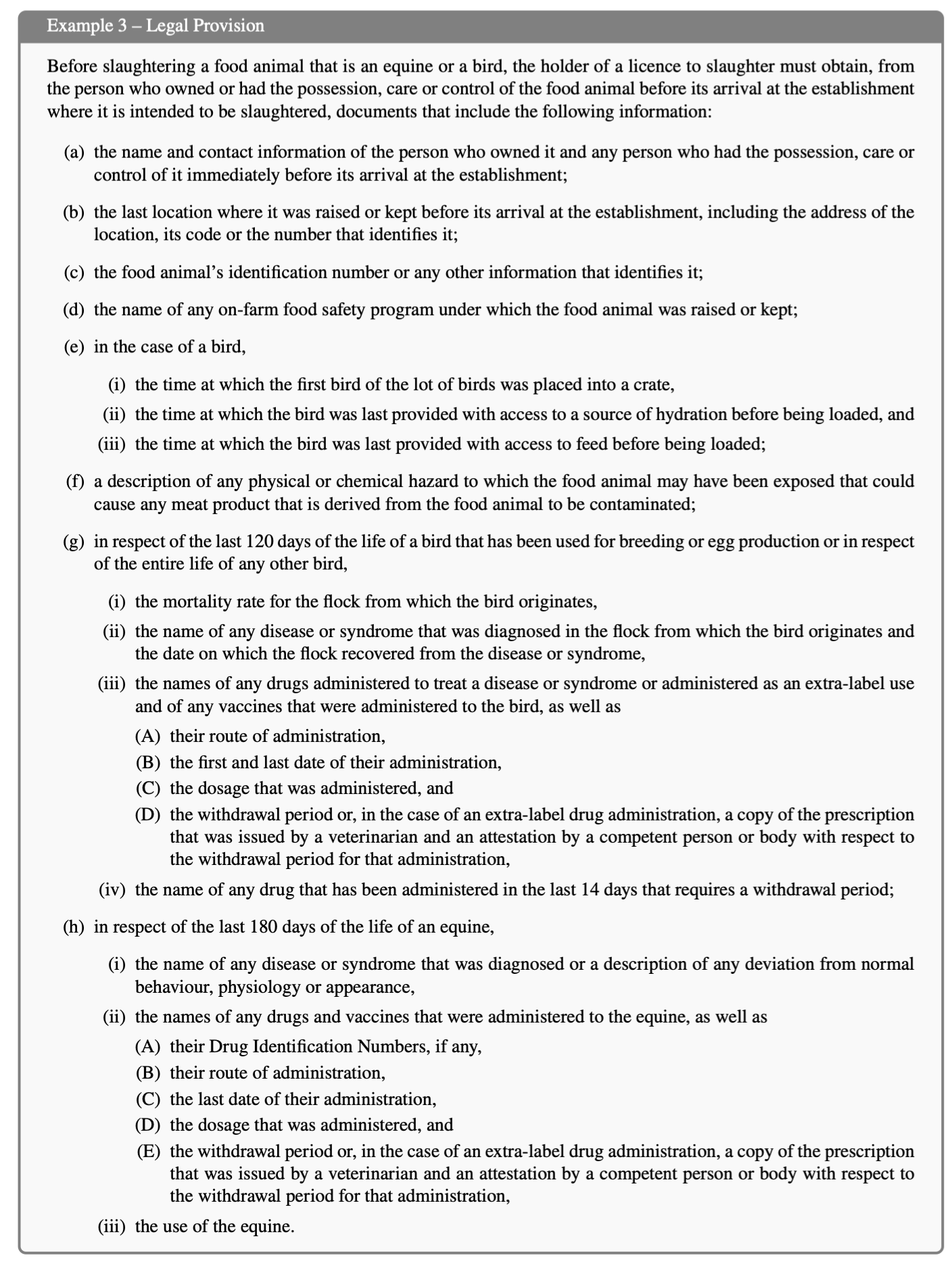}
    \caption{Text of the legal provision on which Example 3 in Section \ref{sec:discussion} is based.}
    \label{fig:example3provision}
\end{figure}

\clearpage

\begin{figure}
  \centering
  \includegraphics[width=\linewidth]{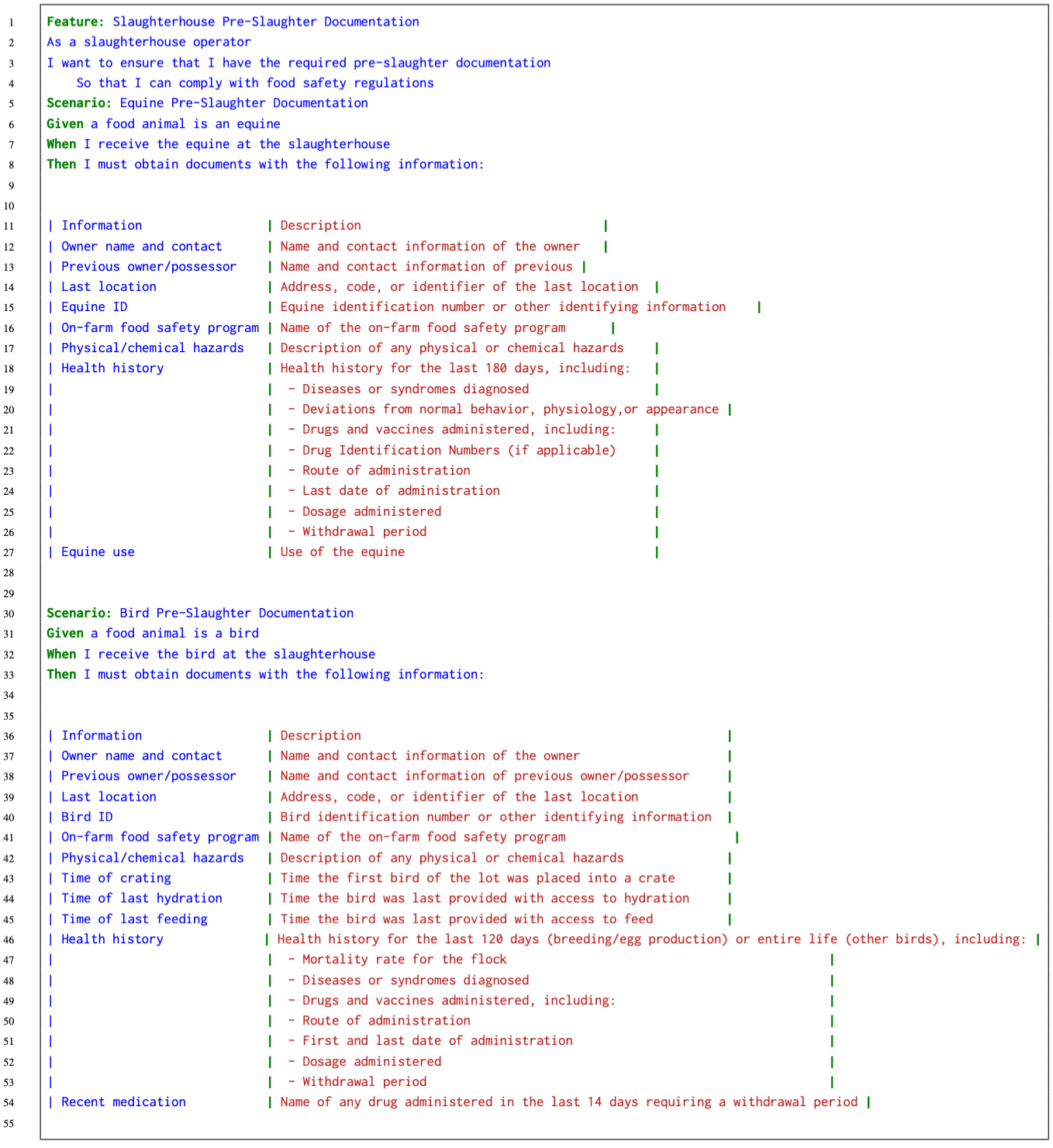}
    \caption{LLM-derived Gherkin specification for Example~3.}\label{fig:example3original}
\end{figure}

\clearpage

\begin{sidewaysfigure}
    \centering
    \includegraphics[width=23cm]{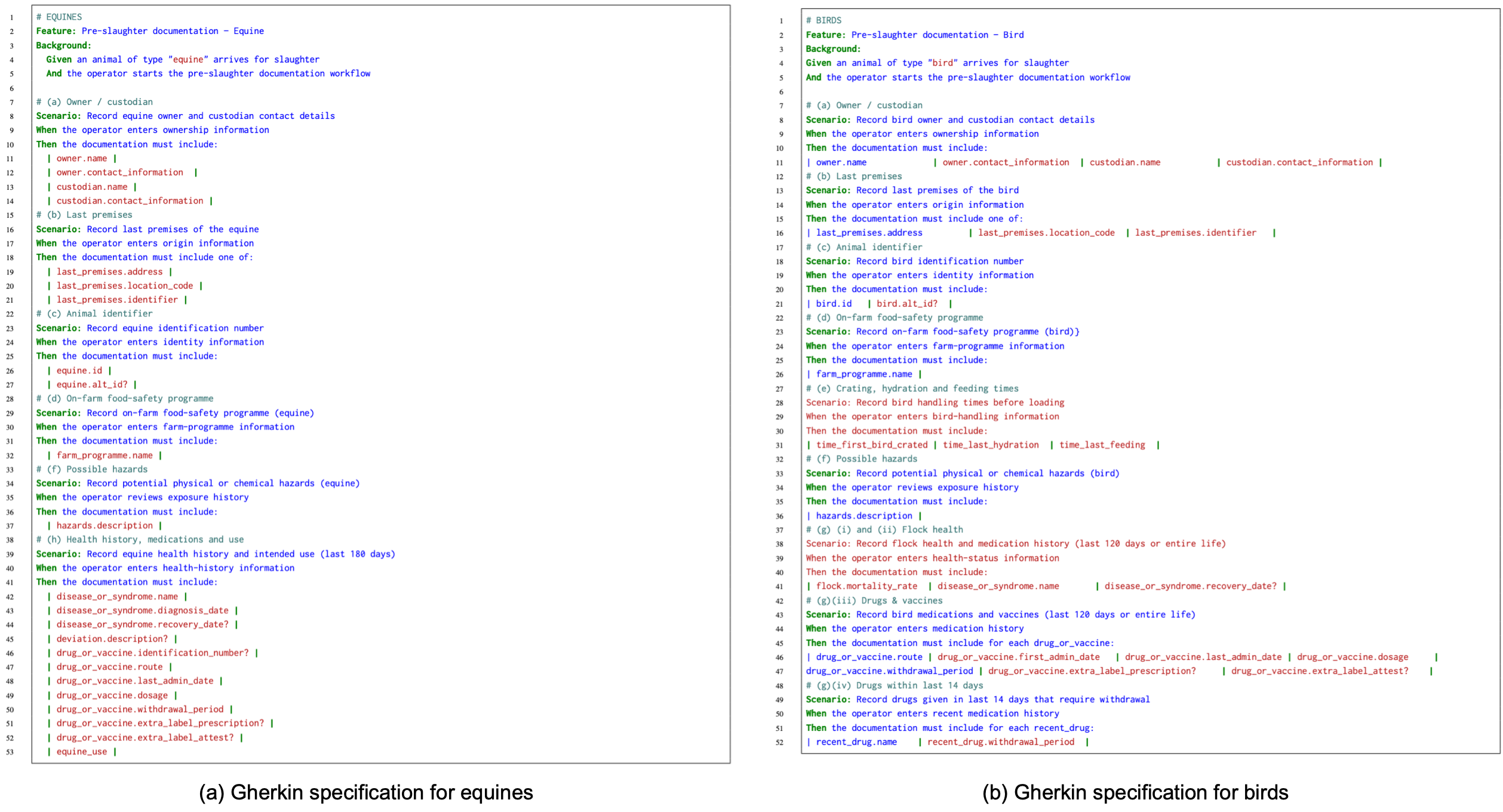}
    \caption{Revised Gherkin specifications for Example 3: (a)~equines and (b)~birds.}
    \label{fig:example3modified}
\end{sidewaysfigure}
\end{document}